\documentstyle[aps,prb]{revtex}
\newtheorem{sn}{}
\renewcommand{\thesn}{%
               \arabic{section}}

\newtheorem{df}{~~~{\sl Definition}}[sn]
\newtheorem{th}[df]{~~~{\bf Theorem}}
\newtheorem{pp}[df]{~~~{\sl Proposition}}
\newtheorem{lm}[df]{~~~{\sl Lemma}}

\newtheorem{ex}{~~~{\sl Example}}

\newcommand{\qed}{$\Box$}

\begin{document}

\title{Characteristic decay of the autocorrelation functions \\ 
prescribed by the Aharonov-Bohm time operator 
}

\author{Manabu Miyamoto 
\thanks{E-mail: miyamo@hep.phys.waseda.ac.jp } 
\\
\sl Department of Physics, Waseda University, Tokyo 169-8555, 
 Japan}
\date{\today}

\maketitle

\begin{abstract} 
The wave functions, the autocorrelation functions of which decay 
faster than $t^{-2}$, for both 
the one-dimensional free particle system and 
the repulsive-potential system 
are examined. It is then shown that such wave functions constitute 
a dense subset of $L^2 ({\bf R}^1)$, 
under several conditions %on potential 
that are particularly satisfied by the square barrier potential 
system. 
It implies that the faster than $t^{-2}$-decay character of 
the autocorrelation function persists against 
the perturbation of potential. 
It is also seen that the denseness of the above subset is 
guaranteed by that of the domain of 
the Aharonov-Bohm time operator. 

\vspace{5mm}


\end{abstract}

%\newpage

\section{Introduction} \label{sec:1}

%max-eqn# is {eqn:1.70}

The time operator $T$ is usually defined 
as an operator that satisfies  
the canonical commutation relation (CCR) 
with the Hamiltonian $H$ for the system considered 
\begin{equation}
[T,H]=i  . 
\label{eqn:1.5}
\end{equation}
This operator has been widely studied in 
the attempt of deriving the time-energy uncertainty relation 
within the %theoretical 
framework of quantum mechanics\ 
(see Ref.\ 1 %\cite{ko} 
and the references therein),  
and of prescribing 
the time-of-arrival in the quantum theory. 
For the latter, 
see a detailed review written by 
Muga and Leavens. \cite{mu}   
In the efforts to solve these problems, 
%there appeared 
a criticism by Pauli\ \cite{pa} 
against the existence of 
the time operator 
has been reexamined. \cite{ga,eg,mi} 
Furthermore, 
there is a possibility that 
the investigation of the time operator  
may reveal its  connection to the quantum dynamics. 
This is expected  firstly  from 
the  definition of the time operator itself, 
i.e. the CCR (\ref{eqn:1.5}),  
which is algebraically so strong  
that their operator characters are mutually restricted. \cite{pu} 
We have a possibility to obtain the information 
of the quantum dynamics through a study of the time operator. 
In this context, 
%From what is mentioned above, 
the investigation of the time 
operator is focused on the commutator 
(not necessarily canonical). 
In particular, it should be remembered that 
%with respect to the spectral property, 
the connection between the commutator of the form 
$[H , iA] =C$ ($C \geq 0$) 
and the spectral property of self-adjoint operators, 
$H, A$ and $C$, has been 
widely studied by 
Putnam \cite{pu} and Kato, \cite{ka} and 
applied to the Schr{\"o}dinger operators by Lavine \cite{la} 
%\cite{la2},  
and others. \cite{re3} 
%We here, however, restrict our consideration to 
%the more strong form,  
%
%
Another implication %symptom
is seen in a statement about 
the ergodic theory in 
classical mechanics,  
which can be formulated in terms of 
%Hilbert space operators, e.g.  
the Liouville operator $L$.    
It states that, for a classical system to be $K$-flow, 
there necessarily exists a self-adjoint operator $Q$ 
satisfying the CCR with the absolutely continuous part of 
$L$: $[Q,L]=i $. \cite{mis,pr} 
Since $L$ and $H$ are mathematically equivalent 
in the role of the generator  of time evolution, 
we may have such a statement in quantum mechanics 
although this expectation  has to %still 
remain in a naive sense. 

In order to examine this possibility, 
a symmetric operator $T$ 
is introduced in Ref.\  6 %\cite{mi} 
on a Hilbert space ${\cal H}$. 
It  satisfies the following operator relation 
with the Hamiltonian operator $H$ on ${\cal H}$: 
\begin{equation} 
T e^{-i tH} = e^{-i tH} (T+t) 
\label{eqn:1.10}
\end{equation}
for all $t \in {\bf R}^1$. 
%, in the sense of the operator. 
Then the CCR\ (\ref{eqn:1.5}) holds 
on the domain ${\rm Dom}(TH)\cap {\rm Dom}(HT)$,   
and so this operator is called the time operator. 
%is derived in the usual manner. 
Mathematically there is no reason why 
we must restrict the relation (\ref{eqn:1.10}) 
to the case of the time operator and the Hamiltonian. 
Actually this relation was investigated in detail by  
Schm{\" u}dgen, who showed that 
the above $T$ and $H$ are unitary equivalent to 
the momentum operator satisfying some boundary condition 
and the position operator, respectively. \cite{sc}  
In the case of the time operator, 
the Aharonov-Bohm time operator $T_0$, \cite{ah} 
(more precisely, its symmetric extension defined later) 
and the free Hamiltonian $H_0$ 
for the one-dimensional free-particle system (1DFPS) 
satisfy the above relation\ (\ref{eqn:1.10}). 
They are defined as, 
\begin{equation}
 T_0  :=  \textstyle{\frac{1}{4}}( QP^{-1}~+~P^{-1} Q ),  ~~~ 
 H_0  :=  P^2 , 
 \label{eqn:1.15}
\end{equation}
where the domain of $T_0$ is defined as 
$ {\rm Dom}(T_0) := 
{\rm Dom}(QP^{-1}) \cap {\rm Dom}(P^{-1} Q ) 
\subset L^2 ({\bf R}^1)$. 
The operators $P$ and $P^{-1}$ 
are the momentum operator on $L^2 ({\bf R}^1)$ 
and  its inverse, where  
$P$ is defined as 
$P:=-i D_x$, $D_x $ being a differential opeartor on 
$L^2 ({\bf R}^1)$. 
Note that $P^{-1}$ is self-adjoint for $P$ is an injection. 
The position operator $Q$ on $L^2 ({\bf R}^1)$ 
is defined as an operator of multiplication 
by $x$. 
Then it is known that $T_0$ is a symmetric operator 
on $L^2 ({\bf R}^1)$. \cite{eg,mi}    
For the later convenience, we here introduce a symmetric 
extension of $T_0$, denoted by $\widetilde{T_0}$, 
which is defined in the momentum representation: 
%the domain of $\widetilde{T_0}$ is  
\begin{equation}
 {\rm Dom}(\widetilde{T_0} ) := \left\{ 
              \psi \in L^2 ({\bf R}^1) ~\left| ~
              \begin{array}{c}
              \widehat{\psi}  \in 
              AC({\bf R}_k ^1 \backslash \{ 0 \} ), ~~
              \displaystyle{  \lim_{k \rightarrow 0} 
              \frac{\widehat{\psi}  (k)}{|k|^{1/2}} =0 }, \\
              \mbox{ and}\\
              \displaystyle{  
              \int_{{\bf R}_k ^1 \backslash \{ 0 \} }
              \left| \frac{d \widehat{\psi}  (k) /k}{d k} 
              + \frac{1}{k}
              \frac{d \widehat{\psi}  (k)}{d k} \right|^2  
              d k < \infty 
              }
              \end{array} 
              \right. 
              \right\} , 
              \label{eqn:1.40}
\end{equation}
%and its action, 
\begin{equation}
  (\widehat{\widetilde{T_0} \psi} )(k) = 
 \frac{i}{4} \left( 
  \frac{d \widehat{\psi}  (k) /k}{d k} + \frac{1}{k}
  \frac{d \widehat{\psi}  (k)}{d k}  \right) , 
  \mbox{ a.e. } k \in {\bf R}_k ^1 \backslash \{ 0 \} ,~~
  \psi \in {\rm Dom}(\widetilde{T_0} ) 
  \label{eqn:1.50}
\end{equation}
where for any $\psi \in L^2 ({\bf R}^1 )$ 
the symbol $\widehat{\psi}$ denotes $F\psi $,  
$F$ being %the unitary operator known as 
the Fourier transform from $L^2 ({\bf R}^1 )$ 
to $L^2 ({\bf R}_k ^1 )$. 
One can see that the operators $\widetilde{T_0}$ and $H_0$ 
satisfy the relation\ (\ref{eqn:1.10}) \cite{mi}, i.e. 
\begin{equation} 
\widetilde{T_0} e^{-i tH_0 } = e^{-i tH_0 } (\widetilde{T_0} +t) 
\label{eqn:1.60}
\end{equation} 
for all $t \in {\bf R}^1$. 
Here our interest is in 
the connection between the time operator and 
the quantum dynamics. 
%, which can be expected from the strong relation\ (\ref{eqn:1.10}). 
Actually, for every $\psi \in {\rm Dom}(T) $ and 
for every $t \in {\bf R}^1 \backslash \{ 0 \}$, 
an inequality 
\begin{equation}
 \left| \left< \psi , e^{-i tH } \psi   \right> \right| ^2 
 \leq 
 \frac{4 \left( \Delta T\right)_{\psi} ^2  \| \psi \| ^2}{t^2} 
% \frac{4 \left( \Delta T\right)_{\psi} ^2  \| \psi \| ^2 }{t^2}
 \label{eqn:1.20}
\end{equation} 
was recently proved in Ref.\ 6 %\cite{mi}  
from the relation\ (\ref{eqn:1.10}). 
Here  
$\left( \Delta T\right)_{\psi}
:= \| (T- \left< \psi , T \psi   \right>) \psi \|$ 
is the standard deviation 
of $T$ %with respect to 
in the state $\psi$, and 
$\left| \left< \psi , e^{-i tH } \psi   \right> \right| ^2 $ 
is the survival probability at time $t$ 
in the initial state $\psi$. 
We call the latter  the autocorrelation function of 
the initial state $\psi$ for the system considered. 
This quantity 
$\left| \left< \psi , e^{-i tH } \psi   \right> \right| ^2 $ 
is considered as an indicator of the 
%``overlap'' 
overlap of the state 
$e^{-i tH } \psi$ at time $t$ with the initial state $\psi$, 
and thus, the name, the autocorrelation function,  
is here appropriate. 
The inequality\ (\ref{eqn:1.20}) is very important 
because the inequality directly connects 
the familiar quantity, 
the autocorrelation function and 
the not-familiar one, $\left( \Delta T\right)_{\psi} $. 
It should be noticed that there are many attempts to relate 
the time-energy uncertainty relation to 
the autocorrelation function\ (see e.g.  Refs.\ 15, 16, and 17
%\cite{bh,gi,pf}
). 
%Hence it may be natural that the relation\ (\ref{eqn:1.20}) exists. 
We can easily see from the inequality 
that $2 \sqrt{2} \left( \Delta T\right)_{\psi} \|\psi \|$ 
gives an upper bound of the half-time 
$\tau_h (\psi) := \sup \left\{ t \geq 0 ~\left|~ 
\left| \left< \psi , e^{-i tH } \psi   \right> \right| ^2 =1/2 
\right. \right\}$. 
Since the above inequality originates only from  
the algebraic relation (\ref{eqn:1.10}), 
we can also obtain, for the operators 
$\widetilde{T_0} $ and $H_0$, 
\begin{equation}
 \left| \left< \psi , e^{-i tH_0 } \psi   \right> \right| ^2 
 \leq 
% 4 \left( \Delta \widetilde{T_0} \right)_{\psi} ^2  
% \| \psi \| ^2 / t^2 
 \frac{4 \left( \Delta \widetilde{T_0} \right)_{\psi} ^2  
 \| \psi \| ^2 }{t^2}
 \label{eqn:1.25}
\end{equation} 
for all 
$\psi \in {\rm Dom}(\widetilde{T_0} ) $ and 
for all $t \in {\bf R}^1 \backslash \{ 0 \}$. 
It is interesting to ask how the inequality (\ref{eqn:1.25}) 
characterizes the decay-dynamics of the autocorrelation 
function for the 1DFPS. With respect to this question, we first see 
that %for any $\psi \in L^2 ({\bf R}_k ^1)$, 
\begin{equation}
\psi \in {\rm Dom}(\widetilde{T_0} ) ~~ \Longrightarrow ~~ 
\exists C >0, \forall t \in {\bf R}^1 \backslash \{ 0 \}, 
 \left| \left< \psi , e^{-i tH_0 } \psi   \right> \right| ^2 
 \leq 
 \frac{C}{t^2}, 
 \label{eqn:1.70}
\end{equation}
however the converse does not hold. 
%The reason for the latter comes from the following example. 
We can easily find a counter example for the latter. 
Consider the wave function 
$\widehat{\psi_1} (k) = (4/\pi)^{1/4} ke^{-k^2 /2} 
\in L^2 ({\bf R}_k ^1)$. Then one can see that 
$ \left| \left< \psi_1 , e^{-i tH_0 } \psi_1   \right> \right| ^2 $ 
decays like $|t|^{-3}$, however 
$\psi_1 \notin {\rm Dom}(\widetilde{T_0} )$ 
from the definition\ (\ref{eqn:1.40}). 
Note that in spite of this example, 
${\rm Dom}(\widetilde{T_0} )$ is dense in $L^2 ({\bf R}^1 )$.

Our aim in the present paper is to answer the question about what  
the faster than $t^{-2}$-decay property of 
the autocorrelation function 
as in (\ref{eqn:1.20}) implies to 
the quantum dynamics, and 
to clarify the role played by 
the time operator or its domain in this respect.  
To this end, 
we attempt to consider %in the present paper 
two one-dimensional systems, i.e.  
a system of a free particle 
and another system of a particle 
with a repulsive potential, 
described by the Hamiltonian $H_1$.  
%satisfying several conditions on the eigenfunctions 
%of $H_1$. 
In particular, we focus on the square-integrable wave functions, 
autocorrelation functions of which decay 
faster than $t^{-2}$ for both of the two systems. %, i.e. 
Then it shall be found, 
under the several conditions on the eigenfunctions 
of $H_1$, 
that the subset of such square-integrable wave functions, 
denoted by   
${\rm C}(H_0 , 2 )  \cap {\rm C}(H_1 , 2 )$,  
is  dense in $L^2 ({\bf R}^1)$. 
The denseness of such a subset  
implies that the faster than $t^{-2}$-decay character of 
the autocorrelation function 
%for the 1DFPS 
is persistent against the perturbation of potentials,  
among which  the square barrier potential is shown to be included. 
These statements are our main results where given 
in Sec.\ \ref{sec:3}. 
We  will also see that the subset 
${\rm C}(H_0 , 2 ) \cap {\rm C}(H_1 , 2 )$
%this dense subset 
includes the dense subspace of 
${\rm Dom}(\widetilde{T_0} )$, and thus 
the denseness of  ${\rm C}(H_0 , 2 )  \cap {\rm C}(H_1 , 2 )$ 
is guaranteed by that of ${\rm Dom}(\widetilde{T_0} )$. 

%This statement  %reformulated in a slightly generalized form 
%is  an application of its slightly generalized form 
%referred to Proposition\  \ref{pp:stability} in the next section 
%which is our main result.  

In the next section, we mention the motivation to 
consider the two systems, as an approach to  
the examination of the inequality (\ref{eqn:1.20}).  
The definitions of the sets  
${\rm C}(H_i , 2 )$ ($i=0, 1$) are also given there. 
Our results mentioned above 
are referred to as Theorem\ \ref{th:stability} and   
Theorem\ \ref{th:stability2} in Sec.\ \ref{sec:3},  
where the latter theorem is applied to 
the square barrier potential system. 
In order to prove these theorems, we also put 
Proposition\  \ref{pp:stability} that is proved 
in Sec. \ref{sec:5}, 
after its preparation in  Sec. \ref{sec:4}. 
In Sec. \ref{sec:4}, the wave operator, the eigenfunction    
expansions, and  the one-dimensional Lippmann-Schwinger equation 
are introduced. 
The concluding remarks are given in Sec.\ \ref{sec:6}. 
In the Appendix, 
the eigenfunctions for the system of square barrier potential 
are shown to  satisfy the conditions  
needed in applying Theorem\  \ref{th:stability2}. 
%The appendix explains in  detail 
%the  example of Theorem\  \ref{th:stability2}. 

\section{
%Physical explanation of 
%Outlook for 
Intersection ${\rm C}(H_0 , 2 )  \cap {\rm C}(H_1 , 2 )$
}

\label{sec:2}

%max-eqn# is {eqn:2.100}

In this section, we mention the reason for 
%our approach of 
considering the two systems, i.e.  
a system of a free particle and another system of 
a particle with a repulsive potential. 
We first introduce the time operators for 
systems with the repulsive potentials $V(x)$ : 
\begin{equation}
0 \leq V(x) \leq \mbox{const.} < \infty, ~~~~~~ 
V \in L^1 ({\bf R}^1). 
\label{eqn:2.10} 
\end{equation}
Then, it follows from Putnam's results \cite{pu-2} 
in the theory of Schr{\" o}dinger operators,  
that for any  potential $V(x)$ satisfying (\ref{eqn:2.10}), 
the Hamiltonian for the potential system defined by 
$H_1 := H_0 + V$ is a self-adjoint operator on $L^2 ({\bf R}^1)$ 
and the wave operators $W_{\pm}$ defined by 
\begin{equation}
%\[
       W_{\pm} :=\mbox{s-}\lim_{t \rightarrow \pm \infty }
                e^{i tH_1} e^{-i tH_0}  
\label{eqn:2.100} 
%\]
\end{equation}
exist and are unitary operators  on $L^2 ({\bf R}^1)$. 
Here "$\mbox{s-}$" stands for strong limit. 
Furthermore they satisfy 
\begin{equation}
H_1 = W_{\pm} H_0 W_{\pm}^* .  
\label{eqn:2.20}
\end{equation}
The time operators $T_{\pm}$ for the potential system 
%specified by 
with the potential $V(x)$ satisfying (\ref{eqn:2.10}) 
are constructed along 
the similar relation to the above:\ \cite{mi}  
\begin{equation}
T_{1,\pm} := W_{\pm} \widetilde{T_0} W_{\pm}^* , 
\label{eqn:2.30}
\end{equation}
where ${\rm Dom}(T_{1,\pm}) = W_{\pm} {\rm Dom}(\widetilde{T_0})$. 
It should be noted that 
the construction of the time operator for 
the potential system in the above sense has been already 
proposed by Le{\'o}n et al. \cite{leon} 
Several properties of such  time operators 
were revealed by them \cite{leon} and Baute et al. \cite{ba} 
Remembering that $\widetilde{T_0}$ and $H_0$ satisfy 
the relation (\ref{eqn:1.60}), and unitarily transforming 
(\ref{eqn:1.60}) by 
$W_{\pm}$ through (\ref{eqn:2.20}) and (\ref{eqn:2.30}), 
we have 
\begin{equation} 
T_{1,\pm} e^{-i tH_1 } = e^{-i tH_1 } (T_{1,\pm} +t), ~~~ 
\forall t \in {\bf R}^1 . 
\label{eqn:2.40}
\end{equation} 
Thus $T_{1,\pm}$ is just regarded as the time operators for 
the repulsive potential system specified by (\ref{eqn:2.10}). 
We also remember that the inequality (\ref{eqn:1.20}) are 
derived only from the algebraic relation (\ref{eqn:1.10}), so that 
\begin{equation}
 \left| \left< \psi , e^{-i tH_1 } \psi   \right> \right| ^2 
 \leq 
% 4 \left( \Delta T_{1,\pm} \right)_{\psi} ^2  
% \| \psi \| ^2 / t^2 
 \frac{4 \left( \Delta T_{1,\pm} \right)_{\psi} ^2  
 \| \psi \| ^2 }{t^2}, ~~~~~ 
\forall \psi \in {\rm Dom}(T_{1,\pm}), ~ 
\forall t \in {\bf R}^1 \backslash \{ 0 \} 
 \label{eqn:2.50}
\end{equation} 
must hold. It is important to notice that the inequalities 
(\ref{eqn:2.50}) mean the existence of a dense subset of 
$L^2 ({\bf R}^1)$, for each of the potential systems, 
that is composed by the wave functions where  
autocorrelation functions decay faster than $t^{-2}$. 

In order to get rid of the restriction of 
the faster than $t^{-2}$-decay dynamics 
to the domain of the time operator as  in (\ref{eqn:1.70}), 
%it is  convenient to 
let us introduce  %following 
a subset of 
$L^2 ({\bf R}^1)$,  
denoted by ${\rm C}(H ,n) $,   %${\rm C}_{\rm ac}(H ,n) $,   
that is defined, for an arbitrary self-adjoint operator $H$ 
on  $L^2 ({\bf R}^1 )$  and a non-negative real number 
$n $, as 
\begin{equation}
 {\rm C} (H, n) := 
 \left\{ \psi \in %P_{\rm ac} (H) 
 L^2 ({\bf R}^1 ) ~\left|~ 
 \exists C > 0,\ \forall t \in {\bf R}^1 \setminus \{ 0\},\ 
 \left| \left< \psi , e^{-i tH } \psi   \right> \right| ^2 
 \leq  \frac{C^2}{|t|^n }
 \right. 
 \right\}. 
 \label{eqn:2.60}
\end{equation}
Clearly we have 
\begin{equation}
 {\rm Dom}(\widetilde{T_0}) \subset {\rm C} (H_0, 2) \subset 
 L^2 ({\bf R}^1 ), 
 \label{eqn:2.70}
\end{equation}
and 
\begin{equation}
 {\rm Dom}(T_{1,\pm}) \subset {\rm C} (H_1, 2) \subset 
 L^2 ({\bf R}^1 ). 
 \label{eqn:2.80}
\end{equation}
Here it should be noticed that 
${\rm C} (H_0, 2) \neq L^2 ({\bf R}^1 )$, 
since, e. g., a Gaussian wave packet  
$\widehat{\psi_0} (k) = (\pi)^{-1/4} e^{-k^2 /2} 
\in L^2 ({\bf R}_k ^1)$,  has  
its autocorrelation function decaying like  
$|t|^{-1}$. 
We also see that ${\rm Dom}(\widetilde{T_0}) \neq {\rm C} (H_0, 2)$ 
for the relation\ (\ref{eqn:1.70}). 
%
%
%
%Analysis of the inclusion relation (\ref{eqn:2.80}) 
%[and (\ref{eqn:2.50})] 
%for 1DFPS is essentially equivalent to 
%that of the inclusion relation (\ref{eqn:2.70}) 
%[and (\ref{eqn:1.25})] for the potential system, 
%because of the unitary transformation 
%used in (\ref{eqn:2.40}). 
%However we should note that this approach does not reflect 
%the fact that the inclusion relation (\ref{eqn:2.80}) is realized 
%for various potential systems satisfying (\ref{eqn:2.10}). 
%%and in principle 
%%one can distinguish those systems. 
%%%%through the measurements of 
%%%some observables e.g. the particle position, and momentum. 
%So 
%

Now we attempt to combine the relations (\ref{eqn:2.70}) 
and (\ref{eqn:2.80}), i.e. 
\begin{equation}
 {\rm Dom}(\widetilde{T_0}) \cap {\rm Dom}(T_{1,\pm}) 
 \subset {\rm C} (H_0, 2) \cap {\rm C} (H_1, 2) \subset 
 L^2 ({\bf R}^1 ). 
 \label{eqn:2.90}
\end{equation}
%and particularly examine whether the intersection 
%${\rm C} (H_0, 2) \cap {\rm C} (H_1, 2)$ 
%is dense in $L^2 ({\bf R}^1 )$. 
One can find that ${\rm C} (H_0, 2) \cap {\rm C} (H_1, 2)$ 
has an interesting character, that is, if an initial state 
belongs to ${\rm C} (H_0, 2) \cap {\rm C} (H_1, 2)$, then  
its autocorrelation function must decay faster than $t^{-2}$, 
irrespectively of the presence of potential. 
One may have a question whether 
such a state can exist. 
In fact, the intersection is not necessarily 
dense in $L^2 ({\bf R}^1 )$, though 
${\rm C} (H_0, 2)$ and  ${\rm C} (H_1, 2)$ are respectively 
dense in $L^2 ({\bf R}^1 )$. 
The primary motivation for the present paper 
is to answer this question and 
the latter can bring us a further understanding of 
the faster than $t^{-2}$-decay dynamics. 
In the following sections, it will be seen that  
${\rm C} (H_0, 2) \cap {\rm C} (H_1, 2)$ is possible to be dense in 
$L^2 ({\bf R}^1 )$, under the several conditions.  
%with (\ref{eqn:2.10}). 

\section{Denseness of 
${\rm C} (H_0, 2) \cap {\rm C} (H_1, 2)$} 
\label{sec:3}

%max-eqn# is {eqn:3.170}

Consider one-dimensional systems 
with the potential of the class: 
\begin{equation} 
\exists \delta > 2, \exists c > 0, \forall x \in {\bf R}^1 ,~
|V(x)| \leq \frac{c}{(1+|x|)^{\delta }} . 
\label{eqn:3.10}
\end{equation} 
The Hamiltonian for the potential system, denoted by  
$H_1$,  
is then defined by $H_1:=H_0 + V$ 
on $L^2 ({\bf R}^1 )$, where $H_0$ is the free Hamiltonian 
on $L^2 ({\bf R}^1 )$. 
For such  potentials, $H_1$ has no positive eigenvalue.\ 
%point spectrum
\cite{si} 
We also see that the above potential is an Agmon potential 
and thus $H_1$ has no singular continuous spectrum. 
\cite{re3-3} 
Our proof of 
the denseness of ${\rm C} (H_0, 2) \cap {\rm C} (H_1, 2)$ 
is based on seeking an explicit example of the subspace which 
guarantees the denseness of 
${\rm C} (H_0, 2) \cap {\rm C} (H_1, 2)$. 
The next proposition provides us with such a subspace.

{\sl 
\begin{pp} \label{pp:stability} {\sl 
: 
Consider a potential system with 
a Hamiltonian $H_1 := H_0 + V$ 
on $L^2 ({\bf R}^1 )$ and 
suppose that the potential $V(x)$ is in the class 
specified by (\ref{eqn:3.10}). 
In addition, assume that, 
%there is a countable set ${\rm N}$ of ${\bf R}_k ^1$,  
%$which includes $0$ and does not accumulate anywhere, 
%such that 
for each $x \in {\bf I}\ (:={\rm supp} V(x))$ fixed, 
the eigenfunctions of $H_1$ denoted by $\varphi_{\pm } (x,k)$ 
in (\ref{eqn:4.70})  in Sec.\ \ref{sec:4} 
belong to $C^1 ({\bf R}_k ^1 \setminus  \{ 0 \})$, 
%is $C^1 $-function of $k$ on ${\bf R}_k ^1 \setminus {\rm N}$, 
%for all fixed $x \in {\bf R}^1$, 
%where ${\rm N}$ is a non-accumulating countable set 
%which includes $0$,  
%set of Lebesgue measure $0$,  
and satisfy the following three conditions: 
\begin{equation} 
   \gamma_{\pm} := 
   \sup_{x \in {\rm I}, k \in {\bf R}_k ^1 \setminus  \{ 0 \} } 
   |\varphi_{\pm } (x,k) | < \infty , 
   \label{eqn:3.103}
\end{equation}
\begin{equation}
   \delta_{\pm} := 
   \sup_{x \in {\rm I}, k \in {\bf R}_k ^1 \setminus  \{ 0 \} } 
   \left| \frac{ \varphi_{\pm } (x,k) }{k} \right| < \infty , 
   \label{eqn:3.105}
\end{equation}
\begin{equation}
   \gamma_{\partial , \pm} := 
   \sup_{x \in {\rm I}, k \in {\bf R}_k ^1 \setminus  \{ 0 \} } 
   \left| \frac{ \partial \varphi_{\pm } (x,k) }{\partial k}
   \right| < \infty , 
   \label{eqn:3.107}
\end{equation}
where ${\rm I}:= {\rm supp} V(x)$. 
Then for any $\psi \in {\rm Dom}(\widetilde{T_0} ) \cap 
{\cal S}({\bf R}^1)$ 
there is some constant $C >0$ 
%which depends only on $V(x)$ and $\psi $,  
such that 
%and satisfies that  
\begin{equation}
       \left| \left< P_{\rm ac} (H_1) \psi , 
       e^{-it H_1} P_{\rm ac} (H_1) \psi \right> \right| \leq 
       \frac{C}{|t|} 
       \label{eqn:3.110}
\end{equation}
for all  $t \in {\bf R}^1 \setminus \{ 0 \}$, 
that is 
\begin{equation}
P_{\rm ac} (H_1) ({\rm Dom}(\widetilde{T_0} ) \cap 
{\cal S}({\bf R}^1) ) \subset {\rm C}(H_1,2) . 
       \label{eqn:3.160}
\end{equation} 
}
\end{pp}
}

%\noindent 
In the above statement, 
${\cal S}({\bf R}^1)$ denotes the subspace of rapidly 
decreasing functions, and 
$P_{\rm ac} (H_1)$ the projection operator 
associated with the absolutely continuous subspace of 
%of $L^2 ({\bf R}^1 )$ 
%with respect to 
$H_1$\ (e.g. see Refs.\  23 and 24 %\cite{re,ka2}
). 
Use of the projection $P_{\rm ac} (H_1)$ in the statement 
comes from the possibility of the existence of bound states of 
the Hamiltonian $H_1$, depending on  the potential 
in the class (\ref{eqn:3.10}). 
Proposition\ \ref{pp:stability} is applicable to 
the system with such a Hamiltonian. 
%The proof %of the above 
%is performed in the following sections. 
Here we remark that 
the above conditions (\ref{eqn:3.103}) and 
(\ref{eqn:3.105}) ensure the existence of 
the following functions $\Phi_{\pm} (k)$:  
\begin{equation} 
\Phi_{\pm} (k) := \left\{ 
                \begin{array}{cc}
                       \delta_{\pm} |k|   &~~( |k| \leq 1 ) \\ 
                       \gamma_{\pm}      & ~~( |k| > 1 )
                \end{array}
        \right. ,  
\label{eqn:3.115}
\end{equation}
satisfying 
\begin{equation}
\Phi_{\pm}/ |k| \in L^2 ({\bf R}_k ^1 ),  
\mbox{~~~and~~~}  \forall k \in  {\bf R}_k ^1 \setminus  \{ 0 \}, ~
\sup_{x \in {\rm I} }  |\varphi_{\pm } (x,k) |  
\leq \Phi_{\pm} (k). 
\label{eqn:3.117}
\end{equation}
%
%\noindent

It should be  noticed that 
under the condition (\ref{eqn:3.10}), 
\begin{equation}
{\rm C} (H_1 , 2) \subset P_{\rm ac} (H_1 ) L^2 ({\bf R}^1 ) 
\label{eqn:3.170}
\end{equation}
holds. 
This follows from the next lemma. 
%For later convenience, 
Before proving it, 
let us remember the results from 
the theory of Schr{\"o}dinger operators 
%the measure theory 
(e.g. see Refs.\ 23 and 24  %\cite{re,ka2}
): 
Suppose that $H$ is a self-adjoint operator on 
a certain Hilbert space, and define the three subspaces as  
${\cal H}_{\rm pp} (H) := 
\{ \psi \in {\cal H} ~|~ \| E( \cdot ) \psi \|^2 
\mbox{ is pure point } \} $, 
${\cal H}_{\rm ac} (H)  := 
\{ \psi \in {\cal H} ~|~ \| E( \cdot ) \psi \|^2 
\mbox{ is absolutely cotinuous } \}$ and 
${\cal H}_{\rm sing} (H)  := 
\{ \psi \in {\cal H} ~|~ \| E( \cdot ) \psi \|^2 
\mbox{ is cotinuous singular} \}$,  
where  
$\{ E(B) ~|~ B \in {\bf B}^1 \}$ is the spectral measure of $H$ 
and ${\bf B}^1$ is the Borel sets %$\sigma$-field 
%which is generated by all open sets 
of ${\bf R}^1$. 
Then these subspaces are closed and orthogonally decompose 
${\cal H}$, i.e. ${\cal H}= {\cal H}_{\rm pp} (H) \oplus 
{\cal H}_{\rm ac} (H) \oplus {\cal H}_{\rm sing} (H) $. 
Furthermore, defining the projection operators $P_{\rm pp}(H)$, 
$P_{\rm ac}(H)$, and $P_{\rm sing}(H)$, 
associated with the closed subspaces ${\cal H}_{\rm pp} (H)$,  
${\cal H}_{\rm ac} (H)$, and  ${\cal H}_{\rm sing} (H)$, 
respectively, 
then these projection operators and the spectral measure $E(B)$ 
are mutually commutable. 

{\sl 
\begin{lm} \label{lm:Hpp0} {\sl 
Suppose that ${\cal H}$ is a Hilbert space and 
$H$ is a self-adjoint operator on ${\cal H}$, then 
${\rm C}(H) \subset P_{\rm c}(H){\cal H}$, where 
%${\rm C}(H) := \bigcup_{n>0} {\rm C}(H,n)$, 
\[
  {\rm C}(H) := 
  \left\{ 
  \psi \in {\cal H} \left| 
  \exists C >0, \exists n >0, 
  \forall t \in {\bf R}^1 \setminus \{ 0 \},  
  \left| \left< \psi, e^{-itH} \psi \right> \right|^2 
  \leq \frac{C^2}{|t|^n} 
  \right. 
  \right\},   
\]
and  $P_{\rm c}(H):=P_{\rm ac}(H) + P_{\rm sing}(H)$. 
%For any potential $V(x)$ of the class (\ref{eqn:3.10}),  
%$P_{\rm ac} (H_1 ) {\rm C} (H_1 , 2) = {\rm C} (H_1 , 2)$. 
}
\end{lm}
}

{\sl Proof} :  
%$L^2 ({\bf R}^1 )$ is orthogonally decomposed as 
%$L^2 ({\bf R}^1 ) = P_{\rm ac} (H_1 ) L^2 ({\bf R}^1 ) \oplus 
%P_{\rm pp} (H_1 ) L^2 ({\bf R}^1 )$, 
%where  $P_{\rm pp} (H_1 )$  is 
%the projection operator 
%associated with the pure point subspace of 
%%of $L^2 ({\bf R}^1 )$ 
%%with respect to 
%$H_1$ (e.g. \cite{re,ka2}), i.e. the subspace spanned by 
%all bound states of $H_1$. 
%For abbreviation, we write 
%$\psi_{\rm c} = P_{\rm c} \psi $ and 
%$\psi_{\rm pp} = P_{\rm pp} \psi $
For notational simplicity, we abbreviate $P_{\rm c} \psi$ to 
$\psi_{\rm c}$, and $P_{\rm pp} \psi$ to $\psi_{\rm pp}$, for all 
$\psi \in {\cal H}$. 
Then, 
by the virtue of 
$E(B) P_{\rm c} = P_{\rm c} E(B) $ and 
$E(B)  P_{\rm pp} = P_{\rm pp} E(B) $,  
we have $\left< \psi, e^{-itH} \psi \right>  
= \left< \psi_{\rm c} , e^{-itH} \psi_{\rm c} \right> 
+\left< \psi_{\rm pp} , e^{-itH} \psi_{\rm pp} \right> $, 
and  the following inequality 
\begin{eqnarray*}
%\begin{equation}
 \left| \left< \psi_{\rm pp} , e^{-itH} \psi_{\rm pp} \right> \right|^2 
 &=&  \left| \left< \psi, e^{-itH} \psi \right> 
              - \left< \psi_{\rm c} , e^{-itH} \psi_{\rm c} \right> 
              \right|^2 \\
 &\leq & 2\left( \left| \left< \psi, e^{-itH} \psi \right> \right|^2 
         + \left| \left< \psi_{\rm c} , e^{-itH} \psi_{\rm c} 
         \right> \right|^2 \right) ~. 
%\end{equation}
\end{eqnarray*}
Let us now consider a particular $\psi \in {\rm C} (H)$ 
($\psi \neq 0$). 
From the definition of ${\rm C} (H)$, 
we can define for this $\psi$ a continuous function $G(t)$ of 
$t \in {\bf R}^1$ : 
\[ G(t):= \left\{ 
                  \begin{array}{cl}
                  \| \psi \|^2 &~~~ 
                  ( |t| \leq ( \| \psi \|^2 /C)^{-2/n} ) \\ 
                  C|t|^{-n/2} &~~~ 
                  ( |t| > ( \| \psi \|^2 /C)^{-2/n} ) \\ 
                  \end{array}
\right. ~, \]
satisfying 
$\left| \left< \psi, e^{-itH} \psi \right> \right| \leq G(t),  
\forall t \in {\bf R}^1 \backslash \{ 0\}$.  
Then  we have 
\[ \left| \left< \psi_{\rm pp} , e^{-itH} \psi_{\rm pp} \right> \right|^2 
 \leq  2 \left( G(t)^2 
         + \left| \left< \psi_{\rm c} , e^{-itH} \psi_{\rm c} 
         \right> \right|^2 \right) ~.\] 
Integrate both sides of this inequality on $[0,T]$ $(T>0)$ and 
divide them by $T$. 
First, we easily see 
\[ \lim_{T \rightarrow \infty} \frac{1}{T} \int_{[0,T]} 
 G(t)^2 dt =0 ~.\]
Secondly notice that  
$\| E (\cdot ) \psi_{\rm c} \|^2 $ is finite and  continuous, 
that is, 
$\| E ({\bf R}^1) \psi_{\rm c} \| 
=\|  \psi_{\rm c} \| < \infty$ 
and $\| E ( A ) \psi_{\rm c} \|^2 =0 $ 
for any countable set $A \in {\bf B}^1$, 
since $\psi_{\rm c} \in P_{\rm c} (H ) {\cal H}$. 
Then by using Wiener's theorem, \cite{re3-4} we see  
\[ \lim_{T \rightarrow \infty} \frac{1}{T} \int_{[0,T]} 
 \left| \left< \psi_{\rm c} , e^{-itH} \psi_{\rm c} 
 \right> \right|^2 dt =0 . \]
Thus, we obtain 
%\begin{equation} 
\[
 \limsup _{T \rightarrow \infty} \frac{1}{T} \int_{[0,T]} 
\left| \left< \psi_{\rm pp} , e^{-itH} \psi_{\rm pp} \right> \right|^2 
 dt =0 ~. \]
%        \label{eqn:Hpp01}
%\end{equation} 
Furthermore we have 
%Now by the use of Wiener's theorem again, we see also that 
%Eq.\ (\ref{eqn:Hpp01}) means 
\begin{eqnarray*} 
 0 & = & \lim_{T \rightarrow \infty} \frac{1}{T} \int_{[0,T]} 
\left| \left< \psi_{\rm pp} , e^{-itH} \psi_{\rm pp} \right> 
\right|^2  dt \\
 & = & \int \int_{ \{ (\lambda , \mu ) \in {\bf R}^2 
       ~|~\lambda =\mu \} } 
        d \left< \psi_{\rm pp} ,E (\lambda) \psi_{\rm pp} \right> 
        d \left< \psi_{\rm pp} ,E (\mu) \psi_{\rm pp} \right> \\ 
 & = & \sum_{\lambda \in {\bf R}^1} \| E (\{ \lambda \} ) 
       \psi_{\rm pp} \|^4 
       \geq  \sum_{\lambda \in A} \| E (\{ \lambda \} ) 
          \psi_{\rm pp} \|^4  ,
\end{eqnarray*} 
where $A$ is any countable set in ${\bf B}^1$. 
Thus $\psi_{\rm pp}$ must belong to $
P_{\rm c} (H ) {\cal H}$. 
This implies that $P_{\rm pp} \psi =\psi_{\rm pp} =0$, i.e. 
$\psi = P_{\rm c} \psi$, $\forall \psi \in {\rm C}(H)$, 
and the proof is completed.  
%for this $\psi$. 
%Therefore %we have that 
%${\rm C} (H , 2) \subset P_{\rm c} (H ) {\cal H}$ and 
%the proof is completed.  
 ~~\hfill \qed

\vspace{5mm}

%\noindent 
Now the relation (\ref{eqn:3.170}) is understood from 
the following argument. 
From the definition of ${\rm C}(H_1, 2)$ in 
(\ref{eqn:2.60}) and the above lemma, we have 
${\rm C}(H_1, 2) \subset P_{\rm c} (H_1 ) L^2 ({\bf R}^1 )$. 
On the other hand, as stated just after (\ref{eqn:3.10}),  
the Hamiltonian $H_1$ defined in Proposition\ \ref{pp:stability} 
has no continuous singular spectrum, i.e. 
$P_{\rm sing} (H_1 ) = 0$, 
and thus we obtain the relation (\ref{eqn:3.170}).

From 
the relations (\ref{eqn:3.160}) and  (\ref{eqn:3.170}),  
we have 
\begin{equation}
P_{\rm ac} (H_1 ) 
({\rm Dom}(\widetilde{T_0} ) \cap {\cal S}({\bf R}^1) )
\subset {\rm C} (H_1 , 2) 
\subset P_{\rm ac} (H_1 ) L^2 ({\bf R}^1 ). 
\label{eqn:3.150}
\end{equation}
Then we see that 
above subsets are densely connected to each other, 
because  ${\rm Dom}(\widetilde{T_0} ) \cap 
{\cal S}({\bf R}^1)$ is dense in $L^2 ({\bf R}^1 )$. 
The latter follows from the existence of 
the dense subspace $F^{-1} {\cal C}_{\rm i}$ 
%\cite{mi} 
satisfying 
$F^{-1} {\cal C}_{\rm i} \subset {\rm Dom}(\widetilde{T_0} ) \cap 
{\cal S}({\bf R}^1)$, 
where  ${\cal C}_{\rm i}$ is defined  in the momentum space as 
%\begin{equation}
\[
{\cal C}_{\rm i} := 
\left\{ 
\left. \widehat{\psi} \in C^{\infty }_0 ({\bf R }_k ^1 ) ~\right| ~
\exists \delta >0 , \forall k \in (-\delta , \delta ), 
\widehat{\psi} (k) = 0 
%\mbox{supp} ~F\psi  \subset {\bf R}_k ^1 \backslash \{ 0\} 
\right\}. 
\]
%It is easily seen that $F^{-1} {\cal C}_{\rm i}$ 
%is invariant under the action of $\widetilde{T_0} $. 
%\label{eqn:Ci} 
%\end{equation}
%
%$\mbox{supp}\ F\psi $ denotes the support of $F\psi $,  i.e. , 
%the closure of $\{ k \in {\bf R }_k ^1 ~|~ (F\psi )(k) \neq 0 \}$. 
%
%\noindent 
%
As a summary, we have from the relations (\ref{eqn:2.70}) 
and (\ref{eqn:3.150}) 

{\bf 
\begin{th} \label{th:stability} {\sl 
: If the conditions in Proposition \ref{pp:stability} are 
satisfied, then 
%the intersection 
$P_{\rm ac} (H_1) {\rm C}(H_0 ,2) \cap {\rm C}(H_1 ,2) $ 
is dense in $P_{\rm ac} (H_1) L^2 ({\bf R}^1 )$. 
}
\end{th}
}

Disregarding the singular continuous part of $H_1$, 
$P_{\rm ac} (H_1) L^2 ({\bf R}^1 )$ is generally considered as 
the subspace  spanned by all scattering states of $H_1$. 
If we assume that $V(x)\geq 0$, i.e. the repulsive type, 
in addition to the potential condition (\ref{eqn:3.10}), 
then $V(x)$ clearly satisfies the condition (\ref{eqn:2.10}). 
In such a case, $H_0$ and $H_1$ is unitary equivalent and thus 
$H_1$ is (spectrally) absolutely continuous, i.e. 
$P_{\rm ac} (H_1)$ is an identity operator on 
$L^2 ({\bf R}^1 )$. This consideration leads us to 
the next theorem, that was announced in Secs.\ 
\ref{sec:1} and \ref{sec:2}.

{\bf 
\begin{th} \label{th:stability2} {\sl 
: If the conditions in Proposition \ref{pp:stability} 
and $V(x)\geq 0$ are satisfied, then 
%the intersection 
${\rm C}(H_0 ,2) \cap {\rm C}(H_1 ,2) $ 
is dense in $L^2 ({\bf R}^1 )$. 
}
\end{th}
}
%}
%
%
%where $P_{\rm ac} (H_1) $ is the projection associated to 
%the absolutely continuous subspace with respect to $H_1$, 
%appears in (\ref{eqn:4.20}). 
%
%

We here mention an application of Proposition\ \ref{pp:stability} 
and Theorem\ \ref{th:stability2} 
to the square barrier potential system. 

{\sl 
\begin{ex} \label{ex:1}  : {\em 
Consider the square barrier potential system 
with the potential: 
\begin{equation} 
V(x) := \left\{ 
                \begin{array}{cc}
                       V_0   &~~( |x| \leq a/2  ) \\ 
                       0      & ~~( |x| > a/2  )
                \end{array}
        \right.  
\label{eqn:3.130}
\end{equation}
where  $a>0$ and $V_0 > 0$. 
The Hamiltonian $H_1$ for the system 
is defined by $H_1:=H_0 + V$ on $L^2 ({\bf R}^1 )$ 
and is self-adjoint.  
Clearly $V(x)$ satisfies the condition (\ref{eqn:3.10}) and 
$V(x) \geq 0$, 
and thus 
%especially the wave operators defined by (\ref{eqn:3.20}) 
%are unitary because 
$P_{\rm ac}(H_1)={\bf 1}$%on $L^2 ({\bf R}^1 )$
, i.e. $H_1$ has no bound states or 
point spectrum. %\cite{pu-2}. 
As shown explicitly in the Appendix, 
the eigenfunctions of $H_1$ 
satisfy the conditions in Proposition\ \ref{pp:stability}, 
%about the differentiability of them about $k$,  and the equations 
(\ref{eqn:3.103}), 
(\ref{eqn:3.105}), and  (\ref{eqn:3.107}). 
Thus Proposition\ \ref{pp:stability} is applicable to this system 
and %we have that 
for any $\psi \in {\rm Dom}(\widetilde{T_0} ) \cap 
{\cal S}({\bf R}^1)$, for any $V_0 >0$, and for any $a >0$ 
there is some constant $C >0$ 
%which depends only on $V_0$, $a$, and $\psi $,  
such that 
%and satisfies that  
\begin{equation}
       \left| \left< \psi , e^{-it H_1} \psi \right> \right| \leq 
       \frac{C}{|t|} 
       \label{eqn:3.140}
\end{equation}
for all  $t \in {\bf R}^1 \setminus \{ 0\} $. 
In particular 
${\rm C}(H_0 ,2)  \cap {\rm C}(H_1 ,2) $ 
is dense in $L^2 ({\bf R}^1 )$. 
}
\end{ex}
}

\section{
Autocorrelation function for potential systems} 
\label{sec:4}

%max-eqn# is {eqn:4.100}

For the preparation of the proof of 
Proposition\ \ref{pp:stability}, 
we shall introduce the wave operator, 
the eigenfunction expansions, and 
the one-dimensional Lippmann-Schwinger equation. 
The wave operators $W_{\pm}$ are defined by 
%\begin{equation}
\[
       W_{\pm} :=\mbox{s-}\lim_{t \rightarrow \pm \infty }
                e^{i tH_1} e^{-i tH_0},   
%\label{eqn:2.100} 
\]
%\end{equation}
where $H_1 := H_0 + V$ with the potential $V(x)$ of the class 
(\ref{eqn:3.10}), different from (\ref{eqn:2.10}). 
%\begin{equation}
%       W_{\pm} :=\mbox{s-}\lim_{t \rightarrow \pm \infty }
%                e^{i tH_1} e^{-i tH_0}  
%\label{eqn:3.20}
%\end{equation}
%where "$\mbox{s-}$" means strong limit 
%and $W_- $ will be used later. 
Since $V(x) \in L^1 ({\bf R}^1 ) \cap L^2 ({\bf R}^1 )$, 
these wave operators %defined for the above Hamiltonian 
exist and are complete. \cite{ku} 
However, in this case, $W_{\pm}$ are generally 
partially-isometric rather than unitary, i.e. 
\begin{equation}
%     \mbox{ and } 
     W_{\pm} ^* W_{\pm} = {\bf 1} ,~~
     W_{\pm} W_{\pm} ^* =  P_{\rm ac} (H_1) , 
\label{eqn:4.20}
\end{equation}
and in particular %for every $t \in {\bf R}^1$
\begin{equation}
     e^{-itH_1} W_{\pm} =  W_{\pm} e^{-itH_0} , 
     \forall t \in {\bf R}^1 . 
%       H_{1, {\rm ac}} = W_{\pm} H_0 W_{\pm} ^* . 
\label{eqn:4.30}
\end{equation}
Then it follows from the use of %the wave operator, 
(\ref{eqn:4.20}) and (\ref{eqn:4.30}) 
%particularly $W_-$,  
that 
\begin{equation}
       \left< P_{\rm ac} (H_1) \psi , 
       e^{-i t H_1} P_{\rm ac} (H_1) \psi \right> 
%       = \left< \psi ,  
%       W_{\pm} e^{-i t H_0} W_{\pm} ^*  \psi \right> 
       = \left< W_{\pm} ^* \psi , 
       e^{-i t H_0} W_{\pm} ^*  \psi \right> 
\label{eqn:4.40}
\end{equation}
for all $\psi \in L^2 ({\bf R}^1 )$. 
By the use of the eigenfunction expansions, 
we can explicitly express  $W_{\pm} ^*  \psi$ 
%is possible to be explicitly represented  
in the momentum representation. 
In particular, for  some fixed numbers $s$ and $s^{\prime}$ 
satisfying 
\begin{equation}
s+s^{\prime} =\delta,~  s > 3/2,~ s^{\prime} >1/2,~ 
{\rm and}~ s^{\prime}  \leq s , 
\label{eqn:4.50}
\end{equation}
%For $\psi \in {\cal S} ({\bf R}^1 )$, 
we have, without resort to the $L^2$-convergence, 
\begin{equation}
       (\widehat{W_{\pm } ^*  \psi })(k)  = \int_{{\bf R}^1 } 
       \overline{ \varphi_{\pm } (x,k) }~ \psi (x) ~d x, ~~~ 
       \forall \psi \in L^{2, s} ({\bf R}^1 ), 
\label{eqn:4.60}         
\end{equation}
%for all $\psi \in L^{2, s} ({\bf R}^1 )$ 
where bar ( $\bar{ }$ ) stands for complex conjugate.   
$L^{2, s} ({\bf R}^1 )$ is a weighted $L^2$-space 
that consists of all $L^2$-functions $\psi$ 
satisfying $\int_{{\bf R}^1 }  
(1+|x|^2 )^s |\psi (x)|^2 dx < \infty $, and then 
we have under the conditions (\ref{eqn:4.50}) 
the inclusion relation 
$L^{2, s} ({\bf R}^1 ) \subset L^1 ({\bf R}^1 )$. 
The functions $\varphi_{\pm } ( x , k)$, 
defined on ${\bf R}^1 \times {\bf R}_k ^1 \setminus \{ 0\} $, 
are the eigenfunctions of $H_1$ 
(more precisely the absolutely continuous part of $H_1$).  
Under the conditions (\ref{eqn:3.10}) and (\ref{eqn:4.50}),  
$\varphi_{\pm } ( x , k)$ 
are guaranteed to be in $L^{2, -s} ({\bf R}^1 )$, 
so that the integral in (\ref{eqn:4.60}) is finite,  
and  satisfy the one-dimensional Lippmann-Schwinger 
equation: They are explicitly given by 
\begin{equation}
       \varphi_{\pm } (x,k) = (2\pi )^{-1/2} e^{i kx}  
       + g_{\pm } (x,k) , 
       \label{eqn:4.70}
\end{equation}
where %$\varphi_0 (x,k) := e^{i kx} / \sqrt{2\pi}$ and 
\begin{equation}
       g_{\pm } (x,k) := \mp \frac{1}{2i |k|} 
                    \int_{{\bf R}^1 } 
                    e^{ \mp i |k| |x-y|} V(y) ~
                    \varphi_{\pm } (y,k) ~d y  
\label{eqn:4.80}
\end{equation}
for all $x \in {\bf R}^1$ and all $k \in {\bf R}_k ^1 
\setminus \{ 0\} $. 
The same conditions ensure 
that $\varphi_{\pm} (x,k) $ 
are ${\rm C}^1 $-functions of $x$ 
for each $k \in {\bf R}_k ^1 
\setminus \{ 0\} $  fixed and  satisfy the time-independent 
Schr{\" o}dinger equation, 
\begin{equation}
       \left( -\frac{d^2}{d x^2} + V(x) \right) 
       \varphi_{\pm } (x,k) 
       = k^2 \varphi_{\pm } (x,k) ,  
       \label{eqn:4.90}
\end{equation}
in the sense of distribution. 
Here we don't prove the eigenfunction expansions 
stated above, however our proof essentially 
follows  the references\ 27 and 28.   %\cite{re3-2,ku2}. 
%our conditions (\ref{eqn:3.10}) and (\ref{eqn:4.50}) 
%are possible to be more relaxed. 
%However it is not needed here and  
%Let us examine whether the autocorrelation function of $\psi$ 
%can decay faster than $t^{-2}$ in $t$. 
%
%Note that for the potential in consideration, 
%$\overline{ g_{\pm } (x,k)} \psi (x)$ is integrable. 
By the use of \ (\ref{eqn:4.60}) and (\ref{eqn:4.70}),  
the probability amplitude 
(\ref{eqn:4.40}) is divided into four terms, 
\begin{equation}
\begin{array}{ll}
        \left< P_{\rm ac} (H_1) \psi , 
        e^{-i t H_1} P_{\rm ac} (H_1) \psi \right>  = & 
       \left< \widehat{\psi}  , e^{-i t k^2} 
       \widehat{\psi}  \right> \\ 
       & \displaystyle{ 
       + \left<  \int_{{\bf R}^1} 
         \overline{ g_{\pm} (y,k) }~\psi (y)~  d y, 
         e^{-i t k^2 } \widehat{\psi} \right> 
       } \\ 
       & % 
       \displaystyle{ 
       + \left<  e^{i t k^2 } \widehat{\psi} , \int_{{\bf R}^1} 
         \overline{ g_{\pm} (y,k) }~\psi (y) ~ d y  \right>   
       } \\ 
       & 
       \displaystyle{  
       + \left<  \int_{{\bf R}^1} 
         \overline{ g_{\pm} (y,k) }~\psi (y) ~ d y , 
         e^{-i t k^2 } 
         \int_{{\bf R}^1} 
         \overline{ g_{\pm} (y,k) }~\psi (y) ~ d y  \right> } ,  
       \label{eqn:4.100}
\end{array}
\end{equation}
where $\psi \in L^{2, s} ({\bf R}^1 )$. 
This expression will be  used 
in the proof of Proposition\ \ref{pp:stability},  
in the next section.

\section{ Proof of Proposition\ 3.1 
%\ref{pp:stability}
%
%Evaluation of equation (9)
%(\ref{eqn:4.100})
} 
\label{sec:5}

\setcounter{df}{0}

Supposing $\psi \in {\rm Dom} (\widetilde{T_0}) \cap 
{\cal S}({\rm R}^1 )$, 
one can find  the following statement, 
which is used in the proof of 
Proposition\ \ref{pp:stability}.

{\sl 
\begin{lm} \label{lm:Dom(T_0)} {\sl : 
If $\psi \in {\rm Dom} (\widetilde{T_0}) \cap 
{\cal S}({\rm R}^1 )$, then $\widehat{\psi} (0) = \widehat{\psi} ^{\prime} (0) =0 $ 
in which $\widehat{\psi} ^{\prime} (k) := d \widehat{\psi} (k) / d k $. 
In particular, 
${\rm Dom}(\widetilde{T_0} ) \cap 
{\cal S}({\bf R}^1) \subset 
{\rm Dom}(QP^{-1} ) \cap {\rm Dom}(P^{-1} Q) 
\subset {\rm Dom}(P^{-2} ) $.   
}
\end{lm}
}

\noindent 
{\sl Proof } : 
Let us prove this in the momentum representation. 
We first note that, for every $\widehat{\psi}  
\in {\cal S}({\rm R}_k ^1 )$, 
$\widehat{\psi}  \in {\rm C}^{\infty} ({\rm R}_k ^1 )$ and 
$\sup_{k \in {{\rm R}_k ^1 } } 
\left| \widehat{\psi} ^{\prime \prime } (k) \right| < \infty $. 
It also follows that $\widehat{\psi} (0) =0$  because of 
$\psi  \in {\rm Dom} (\widetilde{T_0} ) $ 
and the definition\ (\ref{eqn:1.40}). 
Then one can show that for any $k \in (-1 , 1) $, 
there are some numbers $\theta_0 (k)$ and $\theta_1 (k)$ 
such that $0 < \theta_0 (k) <1$, $0< \theta_1 (k) <1 $ and 
\[
  \begin{array}{c}
    \widehat{\psi} (k) = \widehat{\psi} ^{\prime } (0) k + 
    \textstyle{\frac{1}{2}} 
    \widehat{\psi} ^{\prime \prime} ( \theta_0 (k) k ) k^2 , 
    \\ 
    \widehat{\psi} ^{\prime } (k) = \widehat{\psi} ^{\prime } (0) + 
    \widehat{\psi} ^{\prime \prime} ( \theta_1 (k) k ) k .  
  \end{array}
\] 
It follows from the above equations  that 
\[
  \begin{array}{rl}
       (\widehat{\widetilde{T_0} \psi}  ) (k) 
        = & \displaystyle{
            -\frac{i }{2k^2} \widehat{\psi} (k) + \frac{i }{k} 
           \widehat{\psi} ^{\prime } (k) } \\ 
        = & \displaystyle{
            \frac{i}{2k} \widehat{\psi} ^{\prime } (0) 
           -\frac{i }{4} \widehat{\psi} ^{\prime \prime} 
           ( \theta_0 (k) k ) 
           +i \widehat{\psi} ^{\prime \prime} 
           ( \theta_1 (k) k ) , } 
  \end{array}
\]
and thus 
\[
\left| \frac{\widehat{\psi} ^{\prime } (0) }{k} \right| 
\leq 
2 \left| (\widehat{\widetilde{T_0} \psi}  ) (k) \right| + 
\case{5}{2} \sup_{k \in {{\rm R}_k ^1 } } 
\left| \widehat{\psi} ^{\prime \prime } (k) \right| . 
\]
This result implies that 
 $ \widehat{\psi} ^{\prime } (0) = 0 $, 
because $(\widehat{\widetilde{T_0} \psi} ) (k) $ 
is integrable on $(-1, 1)$ for $\widehat{\widetilde{T_0} \psi}  
\in L^2 ({\rm R}_k ^1 )$. 
Thus the first part of the lemma has been proved. 
We now easily see that 
if $\psi \in {\rm Dom} (\widetilde{T_0}) \cap 
{\cal S}({\rm R}^1 )$, then $\widehat{\psi}  \in 
{\rm Dom}(i D_k k^{-1}) \cap 
{\rm Dom}(k^{-1} i D_k ) \subset {\rm Dom}(k^{-2} )$. 
Since 
$i D_k = FQF^{-1} $ and $  k^{-1} = FP^{-1} F^{-1}$,   
this means the latter in the lemma is valid. 
\hfill \qed 

\vspace{5mm}

One can see from (\ref{eqn:1.15}) and this lemma that 
${\rm Dom}(\widetilde{T_0}) \cap {\cal S}({\bf R}^1 ) = 
{\rm Dom}(T_0) \cap {\cal S}({\bf R}^1 ) $. 
Examples of wave functions in 
${\rm Dom}(\widetilde{T_0}) \cap {\cal S}({\bf R}^1 ) $ 
include $\widehat{\phi}_n (k) := k^n e^{-a k^2} $ 
($a>0 $ and integers $n \geq 2$) 
which were used in Ref.\ 1 %\cite{ko}  
by Kobe. 

\vspace{5mm}

\noindent 
{\sl Proof of Proposition\ \ref{pp:stability}} : 
In this proof, we suppose 
$\psi \in {\rm Dom}(\widetilde{T_0} ) \cap 
{\cal S}({\bf R}^1)$ as prescribed in the conditions in 
Proposition\ \ref{pp:stability}. 
Then clearly ${\cal S} ({\bf R}^1 ) \subset L^{2, s} ({\bf R}^1 )$ 
and the expression (\ref{eqn:4.100}) is applicable to the proof. 
Furthermore, since $\psi \in {\rm Dom}(\widetilde{T_0} )$, 
the first term in the right-hand side 
of (\ref{eqn:4.100}) has to decay 
%with an inverse power of $t$ 
faster than $|t|^{-1}$ 
by virtue of the inequality (\ref{eqn:1.25}). 
Therefore, in order to prove the proposition, it is sufficient 
to show that the remaining three terms in the right-hand side 
of (\ref{eqn:4.100})  decay, at least 
like $|t|^{-1}$, or more rapidly.   

We first examine the integral 
$
          \int_{{\bf R}^1} 
         \overline{ g_{\pm} (y,k) }~\psi (y) ~ d y 
$  in (\ref{eqn:4.100}) 
and its derivative 
$
         d \left( \int_{{\bf R}^1} 
         \overline{ g_{\pm} (y,k) }~\psi (y)  d y 
         \right) / d k 
$. 
We  note from Lemma\ \ref{lm:Dom(T_0)} 
that $\psi \in {\rm Dom} ( QP^{-1})$. 
It then follows  that  
$\psi (x)= (P P^{-1} \psi )(x) = -i d ( P^{-1} \psi )(x) /d x ~
{\rm a.e.} ~x $, 
and 
$\lim_{x \rightarrow \pm \infty } ( P^{-1} \psi )(x) = 0 $
for $P^{-1} \psi \in {\rm Dom} ( P)$. 
Moreover $P^{-1} \psi \in L^1 ({\bf R}^1 )$ 
since ${\rm Dom}( Q) \subset L^1 ({\bf R}^1 )$, 
which follows from 
$ 
\int_{{\bf R}^1 } |f(x)| dx \leq 
(
\int_{{\bf R}^1 } |1+|x||^{-2} dx 
\int_{{\bf R}^1 } |1+|x||^{2} |f(x)|^2 dx 
)^{1/2} < \infty,~ 
\forall f \in {\rm Dom}( Q)
$. 
These properties enable us to derive 
%Partial integration used in the above is  
\begin{eqnarray}
        & & 
        \int_{{\bf R}^1} 
        e^{ \pm i |k| |y-x|} \psi (y) 
        ~d y 
        \nonumber \\ 
        &  & 
        ~~~~~ = 
        \int_{ [x, \infty ) } 
        e^{ \pm i |k| (y-x)} \psi (y) 
        ~d y + 
        \int_{ (-\infty , x ]} 
        e^{ \mp i |k| (y-x)} \psi (y) 
        ~d y 
        \nonumber \\ 
        &  & 
        ~~~~~ = 
        \lim_{y\rightarrow \infty } 
        -i e^{ \pm i |k| (y-x)} (P^{-1} \psi )(y) 
        +i (P^{-1} \psi )(x) 
        \mp |k| \int_{ [x, \infty ) } 
        e^{ \pm i |k| (y-x)}  (P^{-1} \psi )(y) 
        ~d y 
        \nonumber \\ 
        & & 
         ~~~~~~~~ 
        -i (P^{-1} \psi )(x) 
        + \lim_{y\rightarrow -\infty } 
        i e^{ \mp i |k| (y-x)} (P^{-1} \psi )(y) 
        \pm |k| \int_{ (-\infty , x ] } 
        e^{ \mp i |k| (y-x)}  (P^{-1} \psi )(y) 
        ~d y 
        \nonumber \\ 
        &  & 
        ~~~~~ = 
        \mp |k| \int_{{\bf R}^1 }
        \frac{y-x}{|y-x|}
        e^{ \pm i |k| |y-x|} 
        (P^{-1} \psi ) (y)  d y  . 
        \label{eqn:5.45} 
\end{eqnarray}
Recalling the assumption (\ref{eqn:3.10}) 
which implies $ \int_{\rm I} |V(x)| dx < \infty $, 
and (\ref{eqn:3.103}), 
we can  obtain from 
Fubini's theorem and  the above result that 
\begin{eqnarray}
          \int_{{\bf R}^1} 
         \overline{ g_{\pm} (y,k) }~\psi (y) ~ d y 
        & = &
        \frac{\pm 1}{2i |k|} \int_{{\rm I}} 
        \left( 
        \int_{{\bf R}^1} 
        e^{ \pm i |k| |y-x|} \psi (y) 
        ~d y 
        \right) 
        V(x) ~ \overline{ \varphi_{\pm } (x,k) } ~d x 
        \nonumber \\ 
        & = &
        \frac{i}{2} \int_{{\rm I}} 
        \left( 
        \int_{{\bf R}^1 }
        \frac{y-x}{|y-x|}
        e^{ \pm i |k| |y-x|} 
        (P^{-1} \psi ) (y)  d y 
        \right) 
        V(x) ~ \overline{ \varphi_{\pm } (x,k) } ~d x , 
         \label{eqn:5.50}
\end{eqnarray}
and thus 
%From (\ref{eqn:5.50}), 
%\begin{eqnarray} 
\begin{equation}
          \left| 
         \int_{{\bf R}^1} 
         \overline{ g_{\pm } (y,k) }~\psi (y) ~ d y 
         \right| 
          \leq   
         {\textstyle \frac{1}{2}} 
         \int_{\rm I} 
         |V(x)| dx 
         \int_{{\bf R}^1} 
         |(P^{-1} \psi ) (y) | ~ d y  
         ~ |\Phi_{\pm } (k)|  
         \label{eqn:5.55}  \\ 
%         & \leq &  
%         \frac{V_0 a}{2} C_{\varphi } ^{\prime } 
%         \int_{{\bf R}^1} 
%         |(P^{-1} \psi ) (y) | ~ d y  
%         ~ A ~ < \infty 
         ~, 
%         \nonumber
\end{equation}
%\end{eqnarray}
where $\Phi_{\pm } (k)$ are defined by (\ref{eqn:3.115}). 
Then, from (\ref{eqn:3.117}) and (\ref{eqn:5.55}),  
we see that 
$        \left(
         \int_{{\bf R}^1} 
         \overline{ g_{\pm} (y,k) }~\psi (y)  d y 
         \right) /  k 
$ 
is a square integrable function of $k$,  
and is  uniformly bounded 
on ${\bf R}_k ^1 \setminus \{ 0\}$. 
We next  examine the following derivative term 
which is well defined  for all $k \in {\bf R}^1 \setminus \{ 0\}$  
under the assumptions %of $V(x)$ and of $ \varphi_{\pm } (x,k) $ 
in the proposition: For $k>0$, 
\begin{eqnarray*}
         & & 
         \frac{d }{d k} %\left(
         \int_{{\bf R}^1} 
         \overline{ g_{\pm} (y,k) }~\psi (y)  d y 
         %\right) 
         \\ 
         & & ~~~~~ =  
         \pm \int_{{\bf R}^1} 
         \frac{i }{2k^2 }
         \left( 
                    \int_{{\rm I} } 
                    e^{ \pm i k |y-x|} V(x) ~
                    \overline{ \varphi_{\pm } (x,k) } ~d x  
         \right) 
          \psi (y)  d y 
         \\ 
         & & 
          ~~~~~~~~   \mp \int_{{\bf R}^1} 
         \frac{i }{2k}
         \left( 
                    \int_{{\rm I} } 
                    \pm i  |y-x| e^{ \pm i k |y-x|} V(x) ~
                    \overline{ \varphi_{\pm} (x,k) } ~d x  
         \right) 
          \psi (y)  d y 
         \\ 
         & & 
          ~~~~~~~~   \mp \int_{{\bf R}^1} 
         \frac{i }{2k}
         \left( 
                    \int_{{\rm I} } 
                    e^{ \pm i k |y-x|} V(x) ~
                    \overline{ 
                    \frac{\partial 
                    \varphi_{\pm} (x,k) }{\partial k} 
                    } ~d x  
         \right) 
          \psi (y)  d y 
          \\ 
         & & 
         ~~~~~ =   
         - \frac{i }{2k }
         \int_{{\rm I} } 
         \left( 
                    \int_{{\bf R}^1} 
                    \frac{y-x}{|y-x|}
                    e^{ \pm i k |y-x|} 
                    (P^{-1} \psi ) (y)  d y 
         \right) 
         V(x) ~
         \overline{ \varphi_{\pm} (x,k) } ~d x  
         \\ 
         & & 
          ~~~~~~~~   + \frac{1}{2k}
         \int_{{\rm I} } 
         \left( 
                    \int_{{\bf R}^1} 
                    |y-x| e^{ \pm i k |y-x|} 
                    \psi (y)  d y 
         \right) 
         V(x) ~ \overline{ \varphi_{\pm} (x,k) } ~d x  
          \\ 
         & & 
          ~~~~~~~~   + \frac{i }{2}
         \int_{{\rm I} }
         \left( 
                    \int_{{\bf R}^1} 
                    \frac{y-x}{|y-x|} 
                    e^{ \pm i k |y-x|} 
                   (P^{-1} \psi ) (y)  d y
         \right) 
         V(x) ~
         \overline{
         \frac{\partial 
         \varphi_{\pm } (x,k) }{\partial k} 
         } ~d x   ~. 
\end{eqnarray*}
In the first equality,  
(\ref{eqn:3.10}), (\ref{eqn:3.107})  
and $\psi \in {\cal S}({\bf R}^1 )$ have been used 
in exchanging the order of differentiation and integration, 
and in the second equality, 
%(\ref{eqn:a.80}),  (\ref{eqn:5.40}),  and 
(\ref{eqn:5.45})  have been used 
at partial integrations. 
It then follows that 
\begin{eqnarray*}
          \left| 
         \frac{d }{d k} 
         \int_{{\bf R}^1} 
         \overline{ g_{\pm } (y,k) }~\psi (y)  d y 
         \right| 
         & \leq & 
         {\textstyle \frac{1 }{2}} 
         \int_{\rm I} 
         |V(x)| dx 
         \int_{{\bf R}^1} 
         |(P^{-1} \psi ) (y) | ~ d y  
         \left| 
         \frac{\Phi_{\pm} (k)}{k } 
         \right| \\ 
         & & 
         + {\textstyle \frac{1 }{2}} 
         \int_{\rm I} 
         |V(x)| dx 
         \int_{{\bf R}^1} 
         |y| | \psi (y) |   d y 
         \left| 
         \frac{\Phi_{\pm} (k)}{k } 
         \right| 
          \\ 
         & & 
         + {\textstyle \frac{1 }{2}} 
         \int_{\rm I} 
         |x| |V(x)| dx 
         \int_{{\bf R}^1} 
         | \psi (y) |   dy 
         \left| 
         \frac{\Phi_{\pm} (k)}{k } 
         \right| 
          \\ 
         &  & 
         + {\textstyle \frac{1 }{2}} 
         \gamma_{\partial \pm } 
         \int_{\rm I} 
         |V(x)| dx 
         \min \left\{ 
         \int_{{\bf R}^1} 
         | \psi  (y) | dy  ~ 
         k^{-1} 
         ,~ 
         \int_{{\bf R}^1} 
         |(P^{-1} \psi ) (y) | d y  
         \right\} ,  
\end{eqnarray*}
for all $k$ where  (\ref{eqn:3.107}), (\ref{eqn:3.117}), and  
$\int_{\rm I } |V(x)| |x| dx < \infty $ have been used. 
The last relation follows from the assumption\ (\ref{eqn:3.10}).  
One can have the similar inequality to the above for $k<0$. 
Therefore $   d \left(
         \int_{{\bf R}^1  } 
         \overline{ g_{\pm} (y,k) }~\psi (y)  d y 
         \right) / d k 
$ 
has to be a square integrable function of $k$ 
from (\ref{eqn:3.117}), and is also seen to be uniformly bounded
on ${\bf R}_k ^1 \setminus \{ 0\}$.

Let us now consider the second and third terms in (\ref{eqn:4.100}). 
Since the third term is essentially equivalent to 
the second, 
%in a way of the time-dependence,   
it suffices to examine the second term in (\ref{eqn:4.100}). 
Then we will show   that 
\begin{equation}
         \left| 
         \left<  \int_{{\bf R}^1} 
         \overline{ g_{\pm} (y,k) }~\psi (y)~  d y, 
         e^{-i t k^2 } \widehat{\psi}  \right> 
         \right| 
         \leq \frac{C_1}{|t|} , 
         \label{eqn:5.30}
\end{equation}
where $C_1 $ is some positive constant  
which depends only on  $\psi$ and the potential.  
The second term in (\ref{eqn:4.100}), 
in particular its $k$-integration over $[0, \infty )$, 
is written by using (\ref{eqn:4.80}) as 
\begin{eqnarray}
         & & 
         \int_0 ^{\infty }  
         \int_{{\bf R}^1} 
         g_{\pm} (y,k) \overline{ \psi (y) } d y\  
         e^{-i t k^2 } \widehat{\psi} (k) dk  
         \nonumber 
         \\ 
        & &  ~~~~~ =  
%        &  = & 
        \frac{i }{2t}          
        \int_0 ^{\infty }  
        \frac{\partial e^{-i t k^2 } }{\partial k} 
        \frac{ \widehat{\psi} (k) }{k} 
        \int_{{\bf R}^1} 
         g_{\pm} (y,k) \overline{ \psi (y) } d y\  
        d k 
         \nonumber 
         \\ 
        & & ~~~~~ =  
%        & = &   
        \frac{i }{2t} 
        \left[ 
        \lim_{k \rightarrow \infty } 
        e^{-i t k^2 } 
        \frac{ \widehat{\psi} (k) }{k} 
        \int_{{\bf R}^1} 
         g_{\pm} (y,k) \overline{ \psi (y) } d y\  
         \right. 
         \nonumber 
         \\ 
         &  & ~~~~~~~~ - 
%        - 
        \lim_{k \downarrow 0 }
        e^{-i t k^2 } 
        \frac{ \widehat{\psi} (k) }{k} 
        \int_{{\bf R}^1} 
         g_{\pm} (y,k) \overline{ \psi (y) } d y\  
         \nonumber 
         \\ 
%        & & 
        & & ~~~~~~~~ 
        - 
        \int_0 ^{\infty } 
        e^{-i t k^2 } 
        \frac{d  \widehat{\psi} (k) /k}{dk} 
        \int_{{\bf R}^1} 
         g_{\pm} (y,k) \overline{ \psi (y) } d y\  
        d k 
         \nonumber 
         \\ 
        & & ~~~~~~~~ 
        \left. 
        - 
        \int_0 ^{\infty }
        e^{-i t k^2 } 
        \frac{ \widehat{\psi} (k) }{k} 
        \frac{d }{dk} 
        \left( 
        \int_{{\bf R}^1} 
         g_{\pm} (y,k) \overline{ \psi (y) } d y  
        \right) 
        d k 
        \right] 
        \nonumber 
        \\ 
        & & ~~~~~ = 
        \frac{-i }{2t} 
        \left[ 
        \int_0 ^{\infty } 
        e^{-i t k^2 } 
        \frac{d  \widehat{\psi} (k) /k}{dk} 
        \int_{{\bf R}^1} 
         g_{\pm} (y,k) \overline{ \psi (y) } d y\  
        d k 
        \right. 
         \nonumber 
         \\ 
        & & ~~~~~~~~ 
        + 
        \left. 
        \int_0 ^{\infty }
        e^{-i t k^2 } 
        \frac{ \widehat{\psi} (k) }{k} 
        \frac{d }{dk} 
        \left( 
        \int_{{\bf R}^1} 
         g_{\pm} (y,k) \overline{ \psi (y) } d y 
        \right) 
        d k 
        \right]
%        \nonumber 
        \label{eqn:5.20}
\end{eqnarray}
where %$\psi \in {\rm Dom}(\widetilde{T_0} ) 
%\cap {\cal S} ({\bf R}^1 )$. 
\[
        \lim_{k \rightarrow \infty } 
        e^{-i t k^2 } 
        \frac{ \widehat{\psi} (k) }{k} 
        \int_{{\bf R}^1} 
         g_{\pm} (y,k) \overline{ \psi (y) } d y  
        =0 \mbox{  and  } 
        \lim_{k \downarrow 0 }
        e^{-i t k^2 } 
        \frac{ \widehat{\psi} (k) }{k} 
        \int_{{\bf R}^1} 
         g_{\pm} (y,k) \overline{ \psi (y) } d y  
         =0  
\]
have been used in the last equality,  
following from  (\ref{eqn:5.55}),  (\ref{eqn:3.115}), 
and the assumption $\psi \in 
{\rm Dom}(\widetilde{T_0} ) \cap {\cal S}({\bf R}^1 ) $. 
Moreover it should be noticed that 
the partial integration used in the second equality 
is justified because both of the two  $k$-integrals in 
(\ref{eqn:5.20}) have finite values.   
To see the latter,  we   note that 
%$\lim_{r \rightarrow 0} \widehat{\psi} (r)/r =0$,  
%the reason for which  
%$\widehat{\psi} (k)/k \in {\rm Dom}(iD_k) 
%\cap {\rm Dom}(k^{-1}) $ from Lemma\ \ref{lm:Dom(T_0)}. 
%Furthermore  
$\widehat{\psi}  /k$ 
and $d \widehat{\psi}  / d k$,   
are square integrable for Lemma\ \ref{lm:Dom(T_0)}, 
and 
$        \left(
         \int_{{\bf R}^1} 
         \overline{ g_- (y,k) }~\psi (y)  d y 
         \right) /  k  
$ 
and 
$   
         d \left(
         \int_{{\bf R}^1 } 
         \overline{ g_- (y,k) }~\psi (y)  d y 
         \right) / d k 
$ are  also square integrable functions of $k$, 
from the results obtained before. 
Therefore we see that  
the two integrals %in the right-hand side of the second equality 
%in (\ref{eqn:5.20}) 
are bounded so that  partial integration %in the second equality 
is well performed. 
These arguments are also applied to 
the $k$-integration over $(-\infty , 0]$, 
and we can obtain (\ref{eqn:5.30}).

We will next  %(shall)
prove that  the last term in the right-hand side of (\ref{eqn:4.100}) 
satisfies 
\begin{equation}
         \left| 
         \left<  \int_{{\bf R}^1} 
         \overline{ g_{\pm } (y,k) }~\psi (y)  d y , e^{-i t k^2 } 
         \int_{{\bf R}^1} 
         \overline{ g_{\pm} (y,k) }~\psi (y)  d y  \right> 
         \right| 
         \leq 
         \frac{C_2}{|t|} ~, 
         \label{eqn:5.35}
\end{equation}
%where $C_2$ is some $t$-independent constant. 
where $C_2 $ is some positive constant  
which depends only on  $\psi$ and the potential.  
In order to see this, we first estimate the following one, 
\begin{eqnarray}
         & &  
         \int_0 ^{\infty }
         \int_{{\bf R}^1} 
         \left| 
         \overline{ g_{\pm} (y,k) } \psi (y)  d y  
         \right|^2 
         e^{-i t k^2 } 
         dk 
          \nonumber 
         \\ 
         & & ~~~~~ = 
         \frac{i}{2 t} 
         \int_0 ^{\infty }
         \left| 
         \int_{{\bf R}^1} 
         \overline{ g_{\pm} (y,k) }~\psi (y)  d y 
         \right|^2  
         \frac{1}{k}
         \frac{ \partial e^{-i t k^2 } }{\partial k }
          d k ~ 
          \nonumber 
         \\ 
         & & ~~~~~ = 
         \frac{i}{2 t} 
         \left[ 
         \lim_{k \rightarrow \infty } 
         \left| 
         \int_{{\bf R}^1} 
         \overline{ g_{\pm} (y,k) }~\psi (y)  d y 
         \right|^2  
         \frac{1}{k}
         e^{-it k^2 } 
         - 
         \lim_{k \downarrow 0 } 
         \left| 
         \int_{{\bf R}^1} 
         \overline{ g_{\pm} (y,k) }~\psi (y)  d y 
         \right|^2  
         \frac{1}{k}
         e^{-i t k^2 } 
         \right]  
         \nonumber 
         \\ 
         & &  ~~~~~~~~ - 
         \frac{i}{2 t} 
         \int_0 ^{\infty }
         \left[ 
         \frac{1}{k } 
         \overline{ \left(
         \int_{{\bf R}^1} 
         \overline{ g_{\pm} (y,k) }~\psi (y)  d y 
         \right) } 
         \frac{d }{d k}\left(
         \int_{{\bf R}^1} 
         \overline{ g_{\pm} (y,k) }~\psi (y)  d y 
         \right) 
         + {\rm c. c. } \right. 
         \nonumber \\ 
         & &  ~~~~~~~~  \left. 
         -\frac{1}{k^2 } 
         \left| 
         \int_{{\bf R}^1} 
         \overline{ g_{\pm} (y,k) }~\psi (y)  d y 
         \right|^2  
         \right] 
         e^{-i t k^2 } 
         ~ d k 
         \nonumber 
         \\ 
         & & ~~~~~ = 
         - 
         \frac{i}{2 t} 
         \int_0 ^{\infty }
         \left[ 
         \frac{1}{k } 
         \overline{ \left(
         \int_{{\bf R}^1} 
         \overline{ g_{\pm} (y,k) }~\psi (y)  d y 
         \right) } 
         \frac{d }{d k}\left(
         \int_{{\bf R}^1} 
         \overline{ g_{\pm} (y,k) }~\psi (y)  d y 
         \right) 
         + {\rm c. c. } \right. 
         \nonumber 
         \\ 
         & &  ~~~~~~~~    \left. 
         -\frac{1}{k^2 } 
         \left| 
         \int_{{\bf R}^1} 
         \overline{ g_{\pm} (y,k) }~\psi (y)  d y 
         \right|^2  
         \right] 
         e^{-i t k^2 } 
         ~ d k ~. 
         \label{eqn:5.70}
\end{eqnarray}
In the last equality, we have used that 
\[
         \lim_{k \rightarrow \infty } 
         \left| 
         \int_{{\bf R}^1} 
         \overline{ g_{\pm} (y,k) }~\psi (y)  d y 
         \right|^2  
         \frac{1}{k}
         e^{-i t k^2 } = 0 ~, ~~~
         \lim_{k \downarrow 0 } 
         \left| 
         \int_{{\bf R}^1} 
         \overline{ g_{\pm} (y,k) }~\psi (y)  d y 
         \right|^2  
         \frac{1}{k}
         e^{-i t k^2 } = 0 ~. 
\]
These follow from (\ref{eqn:5.55}) 
and (\ref{eqn:3.115}). 
We already know that 
$        \left(
         \int_{{\bf R}^1} 
         \overline{ g_{\pm} (y,k) }~\psi (y)  d y 
         \right) /  k 
$ 
and 
$        d \left(
         \int_{{\bf R}^1  } 
         \overline{ g_{\pm} (y,k) }~\psi (y)  d y 
         \right) / d k 
$ 
are square integrable functions of $k$,    
and thus the integrand in (\ref{eqn:5.70}) is 
integrable on ${\bf R}_k ^1$. 
The same argument is also applicable to $k$-integration over 
$(-\infty , 0]$ as in (\ref{eqn:5.70}), and 
hence we obtain (\ref{eqn:5.35}).   
By putting (\ref{eqn:4.100}), (\ref{eqn:5.30}), 
and (\ref{eqn:5.35}) together, we can finally  derive 
(\ref{eqn:3.110}) and the proof 
%of Proposition\ \ref{pp:stability}  
has been completed.  \hfill \qed

%Therefore we have finally reached, 
%by the use of (\ref{eqn:4.100}), (\ref{eqn:5.30}), 
%and (\ref{eqn:5.80}), 
%the expected result that  
%for any $\psi \in {\rm Dom}(\widetilde{T_0} ) \cap 
%{\cal S}({\bf R}^1)$, 
%there is some constant $C >0$, 
%which depends on only $V_0$, $a$, and $\psi $,  
%such that 
%%and satisfies that  
%\begin{equation}
%       \left| \left< \psi , e^{-it H_1} \psi \right> \right| \leq 
%       \frac{C}{|t|} 
%       \label{eqn:5.90}
%\end{equation}
%for all  $t \in {\bf R}^1 $. 

\section{ Concluding remarks 
} 
\label{sec:6}

\setcounter{df}{0}

We would first like to remember Theorem\ \ref{th:stability2}, 
rather than its slightly generalized form, 
Theorem\ \ref{th:stability}, 
because of its clarity for explanation. 
We have considered two one-dimensional systems, i.e. 
the free particle system with the free-Hamiltonian $H_0$ and  
the repulsive-potential system with the Hamiltonian $H_1$, 
and examined the wave functions $\psi \in L^2 ({\bf R}^1 )$ 
where autocorrelation functions 
%for each of the two systems, i.e. 
%$\left| \left< \psi , e^{-it H_i} \psi \right> \right|$ 
%($i=0 \mbox{ or } 1$), 
decay faster than $t^{-2}$ for both systems. 
Theorem\ \ref{th:stability2} states that, 
under the conditions in the statement, 
such  wave functions 
compose a dense subset of $L^2 ({\bf R}^1 )$, 
denoted by 
${\rm C}(H_0 , 2 )  \cap {\rm C}(H_1 , 2 )$. 
The denseness of this subset seems to imply that the 
faster than $t^{-2}$-decay property 
of the autocorrelation function is persistent against  
the perturbation of potential.  
For example, the square-barrier-potential system 
is found to satisfy the conditions in the theorem. 
However these conditions are given in terms of 
the eigenfunctions of the Hamiltonian $H_1$, 
and thus our statement is not straightforward for the practical use.  
Reducing the conditions to explicitly that of the potential 
is a left problem, and needed to know how extensive 
the class of potential  
%against the perturbation of which 
such that 
the faster than $t^{-2}$-decay character remains persistent is. 
Furthermore our conditions is perhaps too strong and 
expected to be relaxed.

Theorem\ \ref{th:stability2} %(and Theorem\ \ref{th:stability}) 
may be thought of as a statement independent of the time operator. 
However, from its derivation, 
we understand that the  existence of 
the Aharonov-Bohm time operator 
$\widetilde{T_0} $ or ${\rm Dom} (\widetilde{T_0} )$ 
guarantees the denseness of 
${\rm C}(H_0 , 2 )  \cap {\rm C}(H_1 , 2 )$ 
in $L^2 ({\bf R}^1 )$. 
Theorem\ \ref{th:stability2} 
can be regarded as a remarkable sign 
%is considered as a crucial evidence 
of the existence of 
the connection between the time operator and 
the quantum dynamics.

%\newpage

\section*{Acknowledgements} 
\setcounter{sn}{1}
\renewcommand{\thesn}{%
          \Alph{sn}}

The author would like to thank Professor I.\ Ohba 
and Professor H.\ Nakazato 
for useful and helpful comments. 
%and the referee for valuable comments. 
He would also like to thank 
%Drs. M.\ Hayasi,\  S.\ Osawa and  
Dr. K.\ Nakazawa 
for useful and helpful discussions.

\appendix

\section*{Eigenfunctions 
for the square barrier potential system 
} 
\label{sec:a}

\setcounter{df}{0}

In this section, we show that eigenfunctions for 
the square barrier potential system satisfy the conditions 
in Proposition\ \ref{pp:stability}, i.e. 
(\ref{eqn:3.103}), (\ref{eqn:3.105}), and (\ref{eqn:3.107}). 
%
%As a preparation for the proof of Proposition\ \ref{pp:stability}, 
We here focus on 
$\varphi_- (x,k) $ in  (\ref{eqn:4.70}),  
however, the same argument is applicable to $\varphi_+ (x,k) $. 
One can see that $\varphi_- (x,k) $ for $k > 0$ is 
the stationary solution for the case where 
a particle approaches from the left of the barrier 
and is reflected or transmitted 
(see, e.g. Ref.\  29 %\cite{LIS}
). 
We define $k_0 := \sqrt{V_0} $, 
and  the support of $V(x)$ as ${\rm I}:=[-a/2, a/2]$ 
for convenience. 
It follows from  (\ref{eqn:4.70}) and  (\ref{eqn:4.90}) 
that for $k > k_0$ 
\begin{equation}
\varphi_- (x,k) = \left\{ 
\begin{array}{ll}
              A e^{i kx} + B(k) e^{-i kx} 
              &  ( x < -a/2 ) \\ 
              C(k) e^{i \kappa x} + D(k) e^{-i \kappa x}
              \qquad &  ( -a/2  \leq  x  \leq a/2 ) \\ 
              F(k) e^{i kx} 
              &  ( a/2 < x ) , 
\end{array}
       \right. 
       \label{eqn:a.10}
\end{equation}
where $A:=(2\pi )^{-1/2}$ and $\kappa :=\sqrt{k^2 -V_0 }$.  
Each coefficient is determined by the property 
stated just before  (\ref{eqn:4.90})   
that 
$\varphi_- (x,k) $ belongs to $C^1 ({\bf R}^1 )$ 
for each $k \in {\bf R}_k ^1 \setminus \{ 0\}$ fixed. 
%or is smooth on the entire line, e.g.  
For $x \in {\rm I}$ and $k > k_0$ we have  
\begin{equation} 
\begin{array}{rl}
       C(k) & =  \displaystyle{ 
       \frac{\kappa +k}{2\kappa } e^{i (k-\kappa )a/2} F(k) 
       }, \\
       D(k) & = \displaystyle{ 
       \frac{\kappa -k}{2\kappa } e^{i (k+\kappa )a/2} F(k) 
       }, \\
       F(k) & =  \displaystyle{ 
                  \left( 
                  \cos \kappa a - 
                  i  \frac{k^2 + \kappa^2 }{2k \kappa } 
                  \sin \kappa a 
                  \right)^{-1} e^{-i ka} A . }
\end{array}
       \label{eqn:a.20}
\end{equation}
In the case $k_0 > k >0 $,  one obtain 
the corresponding equations, through replacing $i \kappa $ 
with $\rho :=\sqrt{V_0 -k^2}$ in (\ref{eqn:a.10}) and 
(\ref{eqn:a.20}). 
At $k=k_0$, we also have 
%from the smoothness condition, that 
\[
\varphi_- (x, k_0 ) = (ik_0 x + 1-ik_0 a/2 ) e^{ik_0 a/2 } 
F(k_0) 
\] 
for all $x \in {\rm I}$ 
where $F(k_0 ) = (1-ik_0 a/2 )^{-1} e^{-ik_0 a } A$. 
Then one can check that $\varphi_- (x, k ) $ is continuous and 
differentiable in $k \in (0, \infty )$, 
and $\varphi_- (x, \cdot ) \in C^1 ({\bf R}_k ^1 \setminus \{ 0\})$, 
for each fixed $x \in {\rm I}$. 
For $k < 0 $, one can substitute -x for x, and -k for k, 
in  (\ref{eqn:a.10}) and 
(\ref{eqn:a.20}) respectively. 
As long as the conditions (\ref{eqn:3.103}), 
(\ref{eqn:3.105}), and (\ref{eqn:3.107}) are concerned,  
it suffices to examine  $\varphi_- (x,k)$ for 
all $x \in {\rm I}$.     
It is noted that for any positive $k$ ( $ \neq k_0 $ ),  
$\left| F(k) \right| \leq A $. Then we have 
\begin{equation}
       \begin{array}{rl}
               | \varphi_- (x,k) |^2 & = \displaystyle{ 
              \left| \cos [ \kappa ( x - a/2 ) ] 
              + i  \frac{k}{\kappa } \sin [ \kappa ( x - a/2 ) ] 
              \right|^2  
              \left| e^{i ka/2} F(k) \right|^2 }   
              \\ 
              & =  \displaystyle{ 
              \left| \cos [ \kappa ( x - a/2 ) ] 
              + i  \left( 
              \frac{\sqrt{k- k_0 } }
              {\sqrt{k+ k_0 } } 
              + \frac{k_0 }{\kappa } 
              \right) 
              \sin [ \kappa ( x - a/2 ) ] 
              \right|^2  
              \left| F(k) \right|^2 } 
              \\ 
              & \leq  
              \left| 
              2 + k_0  | x - a/2 |
              \right|^2  
              |F(k) |^2 \\ 
              & \leq  
              \left| 
              2 + k_0  a
              \right|^2  
              A^2 
              ~,   
       \end{array}
       \label{eqn:a.30}
\end{equation}
for every $x \in {\rm I}$ and $k > k_0 $. 
When $k_0 > k > 0$, we have 
\begin{equation}
       \begin{array}{rl}
               | \varphi_- (x,k) |^2 
              = & \displaystyle{ 
              \left| \cosh [ \rho ( x - a/2 ) ] 
              + i  \frac{k}{ \rho } \sinh [ \rho ( x - a/2 ) ] 
              \right|^2  
              \left| e^{i ka/2} F(k) \right|^2  } 
              \\ 
              = & \displaystyle{ 
              \left| \cosh [ \rho ( x - a/2 ) ] 
              - i  \left( 
              \frac{\sqrt{ k_0 -k} }
              {\sqrt{ k_0 +k} } 
              - \frac{k_0 }{\rho } 
              \right) 
              \sinh [ \rho ( x - a/2 ) ] 
              \right|^2  
              \left| F(k) \right|^2 } 
              \\ 
              \leq & \displaystyle{ 
              \left| \cosh ( k_0 a ) 
              + \sinh ( k_0 a ) 
              + k_0  | x - a/2 | ~
              \frac{\sinh ( k_0 a ) }{k_0 a }
              \right|^2  
              | F(k) |^2  } 
              \\ 
              \leq & 
              \left| \cosh ( k_0 a ) 
              + 2 \sinh ( k_0 a ) 
              \right|^2  
              A^2 ~,   
       \end{array}
       \label{eqn:a.40}
\end{equation}
for every $x \in {\rm I}$. 
As in the same way in the above, 
one can have a similar result for $k < 0$. 
Thus $  \varphi_- (x,k)  $ is   bounded 
on ${\rm I} \times {\bf R}^1 _k \setminus \{ 0\}$ and 
satisfies (\ref{eqn:3.103}), i.e. 
%\begin{equation}
\[
       %C_{\varphi }  := 
       \sup_{ x  \in {\rm I} , k \in {\bf R}^1 _k 
       \setminus  \{ 0 \} 
       %\{ 0 , \pm k_0 \}  
       }  
       | \varphi_- (x,k) | < \infty . 
\]
%       \label{eqn:a.45}
%\end{equation} 
%( {\bf R}_k ^1 \setminus 
%\{ 0 , \pm k_0 \} ) $.  
%for both $x$ and $k$. 
In order to verify the condition\ (\ref{eqn:3.105}), 
we should note that 
\begin{equation}
       \lim_{k \downarrow 0 } \frac{F(k)}{k} 
       = A \left( 
               i \frac{k_0 }{2} \sinh k_0 a 
         \right)^{-1}  ~~~~ {\rm and } ~~~~~
       \lim_{k \rightarrow \infty } \frac{F(k)}{k} =0 ~, 
       \label{eqn:a.70}
\end{equation}
and thus %we have that 
$\sup_{0< k < \infty } |F(k)|/|k| < \infty $,  
because  $F(k)$ is continuous in  
${\bf R}^1 _k \setminus  \{ 0 \} $. 
It follows, from this result and 
(\ref{eqn:a.30}) and (\ref{eqn:a.40}), 
that  $ \varphi_- (x,k)  $ satisfies (\ref{eqn:3.105}). 

Let us next consider the derivative of 
$ \varphi_- (x,k)  $ with respect to $k $. 
When $k > k_0 $, 
it is differentiable at $k $ for each fixed $x \in {\rm I}$,  
and we have 
\begin{equation}
       \begin{array}{rl} 
             \displaystyle{ 
              \frac{ \partial \varphi_- (x,k)  }{\partial k}  }
             = & 
             \displaystyle{ 
             e^{i ka/2} F(k) \frac{ \partial }{\partial k} 
             \left( 
             \cos [\kappa (x-a/2 )] 
             + i  \frac{k}{\kappa } \sin [\kappa (x-a/2 )] 
             \right)  }
             \\ 
             & + \displaystyle{ 
             \frac{\partial e^{i ka/2} F(k)}{\partial k} 
             \left( 
             \cos [\kappa (x-a/2 )] 
             + i  \frac{k}{\kappa } \sin [\kappa (x-a/2 )] 
             \right) , }
       \end{array}
       \label{eqn:a.50}
\end{equation}
in which 
\[
       \begin{array}{l} 
             \displaystyle{ \frac{ \partial }{\partial k} 
             \left( 
             \cos [\kappa (x-a/2 )] 
             + i  \frac{k}{\kappa } \sin [\kappa (x-a/2 )] 
             \right) }
             \\ 
            ~~~~~ = 
            \displaystyle{ -(x-a/2 ) \frac{k}{\kappa } 
            \sin [\kappa (x-a/2 )] 
             +i  \left( 
             \frac{1}{\kappa }
             - \frac{k^2}{\kappa^3 } 
             \right) \sin [\kappa (x-a/2 )]  }
             \\ 
              ~~~~~~~~   \displaystyle{ + i  (x-a/2 ) 
             \frac{k^2}{\kappa^2 } \cos [\kappa (x-a/2 )] }
         \end{array}
%       \label{eqn:a.55}
\]
and 
\[
       \begin{array}{rl} 
        \displaystyle{ \frac{\partial e^{i ka/2} F(k)}{\partial k} } 
        = & 
       \displaystyle{ - \left( 
       - a \frac{k}{\kappa } \sin \kappa a 
       - i  \frac{4k^2 \kappa^2 - (k^2 + \kappa^2 )^2 }
       {2 k^2 \kappa^3 } 
       \sin \kappa a  
       -i a \frac{k^2 + \kappa^2 }{2\kappa^2 } 
       \cos \kappa a \right) }
       \\ 
       & \displaystyle{ \times e^{3i ka/2} F^2 (k) A^{-1}
       - i (a/2)  e^{i ka/2} F(k) . }
       \end{array} 
\]
As for the first term in  (\ref{eqn:a.50}), 
it follows that 
\[ 
   \begin{array}{l}
             \displaystyle{ \left| 
             e^{i ka/2} F(k) \frac{ \partial }{\partial k} 
             \left( 
             \cos [\kappa (x-a/2 )] 
             + i  \frac{k}{\kappa } \sin [\kappa (x-a/2 )] 
             \right) 
             \right| }
             \\
             ~~~~~ \leq  \displaystyle{ |F(k)| \left[ 
                |x-a/2 | (1 + k_0 |x-a/2 | ) 
                +  |x-a/2 | 
                + 2  |x-a/2 | + C_h  |x-a/2 |^3 
                \right] }
             \\ 
             ~~~~~ \leq   (4 + k_0 a + C_h  a^2 ) a A  
    \end{array}
\] 
for all $x \in {\rm I}$ and $k > k_0 $, 
and also for the second term in  (\ref{eqn:a.50}),   
\[    
      \begin{array}{l} 
             \displaystyle{ \left| 
             \left( 
             \cos [\kappa (x-a/2 )] 
             + i  \frac{k}{\kappa } \sin [\kappa (x-a/2 )] 
             \right) 
             \frac{\partial e^{i ka/2} F(k)}{\partial k} 
             \right| }
             \\
             ~~~~~ \leq  \displaystyle{ | 2 + k_0 |x-a/2 | ~| ~ 
                \left[ \left( 
                a (1 + k_0 a ) 
                + 5a  + a^3 V_0 C_h  /2  \right)  |F^2 (k)| A^{-1} 
                + (a/2)|F(k)| 
                \right] }
                \\ 
             ~~~~~ \leq  ( 2 + k_0 a ) ~ 
                \left[ 13/2 
                + k_0 a  
                + a^2 V_0  C_h  /2  
                \right] a A 
        \end{array}
\] 
for all $x \in {\rm I}$ and $k > k_0 $. 
Here we have used the facts 
\[ 
       \left| \frac{k}{\kappa } \sin [\kappa (x-a/2 )] \right| 
        \leq  1 + k_0  |x- a/2 | 
\] 
for every  $k > k_0 $,  
and   $ C_h  :=  \sup_{x > 0 } |h(x)| $, 
where $h(x) := (\sin x -x \cos x )/ x^3 $. 
It should be noted that $C_h < \infty $, 
because $\lim_{x \downarrow 0} h(x) = \case{1}{3}$,  
$\lim_{x \rightarrow \infty } h(x) =0 $ 
and  $h(x)$ is continuous at each $x >0$. 
Hence it can be seen, from 
the above results,  
that $\partial \varphi_- (x,k) / \partial k$ is  bounded on 
${\rm I} \times (k_0 , \infty )$. 
If $k_0 > k >  0 $, 
$\varphi_- (x,k)$ is differentiable with respect to  $k$  
for each fixed $x \in {\rm I}$ and we have  
\begin{equation}
  \begin{array}{rl} 
        \displaystyle{ 
        \frac{ \partial \varphi_- (x,k)  }{\partial k} }
             = & 
             \displaystyle{ 
             e^{i ka/2} F(k) \frac{ \partial }{\partial k} 
             \left( 
             \cosh [\rho (x-a/2 )] 
             + i  \frac{k}{\rho } \sinh [\rho (x-a/2 )] 
             \right)  }
             \\ 
             &  \displaystyle{ + 
             \frac{\partial e^{i ka/2} F(k)}{\partial k} 
             \left( 
             \cosh [\rho (x-a/2 )] 
             + i  \frac{k}{\rho } \sinh [\rho (x-a/2 )] 
             \right) , }
   \end{array}
       \label{eqn:a.60}
\end{equation}
where
\[ 
   \begin{array}{l}
             \displaystyle{ \frac{ \partial }{\partial k} 
             \left( 
             \cosh [\rho (x-a/2 )] 
             + i  \frac{k}{\rho } \sinh [\rho (x-a/2 )] 
             \right) }
             \\ 
            ~~~~~ = \displaystyle{  -(x-a/2 ) \frac{k}{\rho } 
            \sinh [\rho (x-a/2 )] 
             +i  \left( 
             \frac{1}{\rho }
             + \frac{k^2}{\rho^3 } 
             \right) \sinh [\rho (x-a/2 )]  }
             \\ 
              ~~~~~~~~   \displaystyle{ - i  (x-a/2 ) 
             \frac{k^2}{\rho^2 } \cosh [\rho (x-a/2 )] }
%       \label{eqn:a.55}
   \end{array}
\] 
and 
\[ 
   \begin{array}{rl}
        \displaystyle{ 
        \frac{\partial e^{i ka/2} F(k)}{\partial k} } = 
        &  
       \displaystyle{ - \left( 
       - a \frac{k}{\rho } \sinh \rho a 
       - i  \frac{4k^2  \rho^2 + (k^2 - \rho^2 )^2 }{2 k^2 \rho^3 } 
       \sinh \rho a  
       + i a \frac{k^2 - \rho^2 }{2\rho^2 } 
       \cosh \rho a \right) }
       \\ 
       & 
       \displaystyle{ \times e^{3i ka/2} F^2 (k) A^{-1}
       - i (a/2)  e^{i ka/2} F(k) . }
   \end{array}
\]
Concerning the first term in the right-hand side of (\ref{eqn:a.60}), 
we obtain,  for $k_0 > k > 0 $, 
\[ 
   \begin{array}{l}
        \displaystyle{  \left| 
             e^{i ka/2} F(k) \frac{ \partial }{\partial k} 
             \left( 
             \cosh [\rho (x-a/2 )] 
             + i  \frac{k}{\rho } \sinh [\rho (x-a/2 )] 
             \right)  
       \right| }
       \\
       ~~~~~ \leq  
       \displaystyle{ |F(k)| \left[ 
       2 |x-a/2 | \sinh (k_0 |x-a/2| ) 
       + \frac{\sinh (k_0 |x-a/2| ) }{k_0 } 
       \right. }
       \\  
        ~~~~~~~~   \displaystyle{ \left. 
        + \left( |x-a/2 |^3 V_0 C_l 
       + \frac{\sinh (k_0 |x-a/2| ) }{k_0 } 
       + |x-a/2 | \cosh (k_0 |x-a/2| ) \right) 
       \right] }
       \\ 
       ~~~~~ \leq  
       \displaystyle{ \left[ 
       a
       \left( 
       2  \sinh (k_0 a ) + \cosh (k_0 a) 
       +V_0 a^2 C_l 
       \right) 
       + 2 \frac{\sinh (k_0 a ) }{k_0 } 
       \right] A ~, }
   \end{array}
\] 
for all $x \in {\rm I}$, 
and also for the second term in (\ref{eqn:a.60}) 
\[
  \begin{array}{l}
        \displaystyle{ \left| 
              \left( 
             \cosh [\rho (x-a/2 )] 
             + i  \frac{k}{\rho } \sinh [\rho (x-a/2 )] 
             \right) 
             \frac{\partial e^{i ka/2} F(k)}{\partial k} 
       \right| }
       \\ 
       ~~~~~ \leq  
       \displaystyle{ \left( \cosh (k_0 |x-a/2| ) 
         + 2 \sinh (k_0 |x-a/2| ) 
       \right)  \left[ 
       \left( 
         a \sinh (k_0 a ) 
         + \frac{\sinh (k_0 a ) }{k_0 } 
         \right.  \right.  }
       \\ 
        ~~~~~~~~   \displaystyle{  \left. \left. 
       + V_0 a^3 C_l 
       + k_0 \frac{\sinh (k_0 a ) }{2k^2 } 
       \right) |F(k)|^2 A^{-1} 
       + (a/2) |F(k)| 
       \right] }
       \\ 
       ~~~~~ \leq  
       \displaystyle{ \left( \cosh (k_0 a ) 
         + 2 \sinh (k_0 a ) 
       \right) \left[ 
       \left( 
         a \sinh (k_0 a ) 
         + \frac{\sinh (k_0 a ) }{k_0 } 
         + V_0 a^3 C_l 
       \right) A \right. }
       \\ 
        ~~~~~~~~   \displaystyle{ \left. 
       + k_0 \sinh (k_0 a ) (2A)^{-1} 
         \sup_{0< k < k_0 } \frac{|F(k)|^2 }{k^2 } 
       + (a/2) A 
       \right] .      }
  \end{array}
\]
Here we have used the fact that 
$(\sinh x )/x $ is a monotonically increasing function of $x$ 
for $x>0$ and an inequality 
\[
  \left| 
  \frac{k}{\rho} \sinh [ \rho (x-a/2 ) ] 
  \right| 
  \leq 2 \sinh (k_0 |x-a/2 |  ) ~. 
\]
Furthermore, we have defined $C_l :=\sup_{0< x < k_0 a} 
l(x)$, where $l(x):= (\sinh x -x \cosh x)/ x^3 $. Then 
$C_l < \infty $ %is bounded 
because of the relation 
$\lim_{x \downarrow 0} l(x) =-\case{1}{3}$ and 
the continuity of $l(x)$ 
on $(0, \infty )$.  
%for every $x >0$. 
Thus,  
$\partial \varphi_- (x,k) / \partial k$ has to be bounded 
on ${\rm I} \times (0 , k_0 )$. 
See (\ref{eqn:a.70}). 
Hence we have obtained that 
$\varphi_- (x,k)$ is differentiable in 
${\bf R}^1 _k \setminus \{ 0\}$ of $k$ and 
its derivative with respect to $k$ 
is  bounded on  ${\rm I} \times (0 , \infty )$. 
A similar result can be also obtained 
for ${\rm I} \times ( -\infty , 0 )$, and thus 
we have  finally shown  (\ref{eqn:3.107}), i.e.   
%\begin{equation}
\[
       %C_{\partial \varphi }  := 
       \sup_{ x  \in {\rm I} , k \in {\bf R}^1 _k 
       \setminus  \{ 0 \} }
       %\{ 0 , \pm k_0 \}  }  
       \left| \frac{\partial \varphi_- (x,k) }{\partial k} 
       \right| < \infty ~ . 
\]
%       \label{eqn:a.80}
%\end{equation} 

%\clearpage

%\appendix
%\section{List of macros for formatting text, figures and tables}

\end{document}